\documentclass[a4paper,12pt]{article}
\usepackage{amsmath,amssymb,amsthm}

\newtheorem{lemma}{Lemma}
\newtheorem{theorem}{Theorem}

\theoremstyle{definition}

\theoremstyle{remark}

\newcommand{\beq}{\begin{eqnarray}}
\newcommand{\eeq}{\end{eqnarray}}
\newcommand{\beqnn}{\begin{eqnarray*}}
\newcommand{\eeqnn}{\end{eqnarray*}}
\newcommand{\rd}{\partial}

\newcommand{\CC}{\mathbf{C}}
\newcommand{\PP}{\mathbf{P}}

\newcommand{\bslambda}{\boldsymbol{\lambda}}
\newcommand{\bst}{\boldsymbol{t}}
\newcommand{\calS}{\mathcal{S}}

\begin{document}

\title{L\"owner equations, Hirota equations\\
and reductions of universal Whitham hierarchy}
\author{Kanehisa Takasaki$^1$%
\thanks{E-mail: E-mail: takasaki@math.h.kyoto-u.ac.jp}
~and Takashi Takebe$^2$%
\thanks{E-mail: takebe@math.ocha.ac.jp}\\
\\
{\small
$^1$Graduate School of Human and Environmental Studies,
Kyoto University}\\
{\small Yoshida, Sakyo, Kyoto, 606-8501, Japan}\\
{\small $^2$Department of Mathematics, Ochanomizu University}\\
{\small Otsuka 2-1-1, Bunkyo-ku, Tokyo, 112-8610, Japan}}
\date{}
\maketitle

\begin{abstract}
This paper reconsiders finite variable reductions of 
the universal Whitham hierarchy of genus zero in the perspective 
of dispersionless Hirota equations.  In the case of one-variable 
reduction, dispersionless Hirota equations turn out to be 
a powerful tool for understanding the mechanism of reduction.  
All relevant equations describing the reduction (L\"owner-type 
equations and diagonal hydrodynamic equations) can be thereby 
derived and justified in a unified manner.  The case of 
multi-variable reductions is not so straightforward.  
Nevertheless, the reduction procedure can be formulated 
in a general form, and justified with the aid of dispersionless 
Hirota equations.  As an application, previous results of 
Guil, Ma\~{n}as and Mart\'{\i}nez Alonso are reconfirmed 
in this formulation.  
\end{abstract}
\bigskip

\begin{flushleft}
\end{flushleft}
\newpage

\section{Introduction}

The L\"owner equation \cite{Loewner23} 
is a differential equation that describes a family 
of deformations (with parameter $\lambda$) of 
univalent conformal maps $g_\lambda: D \to B_\lambda$ 
from a fixed disk $D \subset \CC\PP^1$ 
to a simply connected domain $B_\lambda 
= B\setminus\Gamma_\lambda \subset \CC\PP^1$ 
with a slit formed by a continuously growing arc 
$\Gamma_\lambda$ on a fixed curve $\Gamma$. 
$g_\lambda$ is the inverse of a univalent conformal map 
$f_\lambda: B_\lambda \to D$ whose existence 
is ensured by Riemann's mapping theorem.  
For technical reasons, we assume that 
$D$ is centered at $\infty$, $B$ contains $\infty$ 
in its interior, and $g_\lambda(\infty) = \infty$. 
To write the L\"owner equation, let us express 
the maps $g_\lambda$ and $f_\lambda$ in terms of 
coordinates as $z = g(p,\lambda)$ and $p = f(z,\lambda)$. 
The L\"owner equation thereby reads 
\beqnn
  \frac{\rd f(z,\lambda)}{\rd\lambda}
  = f(z,\lambda)
    \frac{\kappa(\lambda)+f(z,\lambda)}{\kappa(\lambda)-f(z,\lambda)}
    \frac{\rd\phi(\lambda)}{\rd\lambda} 
\eeqnn
for $f(z,\lambda)$ and 
\beqnn
  \frac{\rd g(p,\lambda)}{\rd\lambda}
  = p\frac{p+\kappa(\lambda)}{p-\kappa(\lambda)}
    \frac{\rd g(p,\lambda)}{\rd p} 
    \frac{\rd\phi(\lambda)}{\rd\lambda}
\eeqnn
for $g(p,\lambda)$. (As one can easily see, 
these two equations are equivalent.)  
$\kappa(\lambda)$ and $\phi(\lambda)$ are 
auxiliary functions that are determined by 
(and, conversely, determine) the deformations 
of the maps.  $\kappa(\lambda)$ is called 
a ``driving force'', and $\phi(\lambda)$ is related 
to the behavior of $g(p,\lambda)$ and $f(z,\lambda)$ 
at infinity: 
\beqnn
\begin{aligned}
  g(p,\lambda) &= e^{\phi(\lambda)}p + O(1) \quad (p \to\infty), \\
  f(z,\lambda) &= e^{-\phi(\lambda)}z + O(1) \quad (z \to\infty). 
\end{aligned}
\eeqnn

A few variant of the L\"owner equation 
are also known.  Firstly, one can consider 
a family of univalent conformal maps $g_{\bslambda}$ 
to a domain $B_{\bslambda}$ with multiple slits of arcs 
on several fixed curves $\Gamma_1,\ldots,\Gamma_M$.  
Since the arcs can grow independently, this family 
depends on $M$-dimensional parameters $\bslambda 
= (\lambda_1,\ldots,\lambda_M)$.  
The univalent homomorphic functions 
$g(p,\bslambda)$ and $f(z,\bslambda)$ representing 
$g_{\bslambda}$ and its inverse $f_{\bslambda}$ 
satisfy a system of L\"owner-like equations 
\beqnn
\begin{aligned}
  \frac{\rd f(z,\bslambda)}{\rd\lambda_j}
  &= f(z,\bslambda)
     \frac{\kappa_j(\bslambda)+f(z,\bslambda)}
          {\kappa_j(\bslambda)-f(z,\bslambda)}
    \frac{\rd\phi(\bslambda)}{\rd\lambda_j}, \\
  \frac{\rd g(p,\bslambda)}{\rd\lambda_j}
  &= p\frac{p+\kappa_j(\bslambda)}{p-\kappa_j(\bslambda)}
     \frac{\rd g(p,\bslambda)}{\rd p} 
     \frac{\rd\phi(\bslambda)}{\rd\lambda_j}, 
  \quad j = 1,\ldots,M, 
\end{aligned}
\eeqnn
with respect to $\lambda_j$'s.   
Secondly, one can choose the upper half plane $H$ 
rather than the disk, and consider univalent 
conformal maps $g_\lambda$ from $H$ to a simply connected 
domain $B_\lambda$ with $\infty$ on its boundary.  
The holomorphic functions $g(p,\bslambda)$ 
and $f(z,\bslambda)$ representing these maps 
satisfy a pair of differential equations 
called the ``chordal L\"owner equation'' (see below).  
This equation, too, have multi-slit analogues. 

Remarkably, these L\"owner-type equations 
also emerge in finite variable (or finite field) 
reductions of various dispersionless integrable systems. 
This fact was first discovered by Gibbons and Tsarev 
in the case of the Benney equations \cite{GT96,GT99}.  
$g(p,\bslambda)$ shows up therein as 
a generating function $g(p)$ of the Benney moments, 
and the equations of motion of those moments 
are converted to an evolution equation for $g(p)$.  
This evolution equation can be identified 
with the Lax equation of the second flow of 
the dispersionless KP hierarchy.  
Gibbons and Tsarev observed that 
a general $M$-variable reduction 
of this system can be obtained in the following form: 
\begin{itemize}
\item[1.] $g(p)$ depends on the spacetime variable 
$(x,t)$ of the Benney equations via $M$ reduced 
variables $\bslambda = (\lambda_1,\ldots,\lambda_M)$ as 
\beqnn
  g(p) = g(p,\bslambda), \quad 
  \bslambda = \bslambda(x,t). 
\eeqnn
\item[2.] $g(p,\bslambda)$ satisfies 
the chordal L\"owner equations 
\beqnn
  \frac{\rd g(p,\bslambda)}{\rd\lambda_j}
  = \frac{1}{p - U_j(\bslambda)}
    \frac{\rd g(p,\bslambda)}{\rd p}
    \frac{\rd a(\bslambda)}{\rd\lambda_j}, 
  \quad j = 1,\ldots,M, 
\eeqnn
of domains $B_{\bslambda}$ with $M$ slits. 
The inverse function $f(z,\bslambda)$ of 
$g(p,\bslambda)$ satisfies the dual equations
\beqnn
  \frac{\rd f(z,\bslambda)}{\rd\lambda_j}
  = \frac{1}{U_j(\bslambda) - f(z,\bslambda)}
    \frac{\rd a(\bslambda)}{\rd\lambda_j}, 
  \quad j = 1,\ldots,M. 
\eeqnn
$U_j(\bslambda)$ and $a(\bslambda)$ are 
auxiliary functions that determine the reduction. 
$U_j(\bslambda)$'s are called driving forces, 
and $a(\bslambda)$ is related to the asymptotic 
behavior of $g(p,\bslambda)$ and $f(z,\bslambda)$ 
at infinity. 
\item[3.] $\lambda_j$'s satisfy a system of 
``diagonal'' hydrodynamic equations 
\beqnn
  \frac{\rd\lambda_j}{\rd t}
  = \chi_j(\bslambda)\frac{\rd\lambda_j}{\rd x}, 
  \quad j = 1,\ldots,M. 
\eeqnn
Thus $\lambda_j$'s are ``Riemann invariants'' 
of the reduced system.  The ``characteristic speeds'' 
$\chi_j(\bslambda)$ turn out to satisfy a set 
of conditions, which enables one to solve 
these equations by Tsarev's generalized 
hodograph method \cite{Tsarev85}.  
\end{itemize}
This result has been generalized to many other cases 
\cite{YG00,BG03,BG04,MMAM02,GMMA03,Manas04,TTZ06}.  
Recently this issue is also studied from the point 
of view of Hamiltonian structures \cite{GLR08}. 

In this paper, we address the problem of 
finite variable reductions of the universal 
Whitham hierarchy of genus zero \cite{Krichever94}.  
This issue was studied by Guil, Ma\~{n}as 
and Mart\'{\i}nez Alonso \cite{GMMA03}, 
and the relevance of L\"owner-type equations 
was already recognized therein.  We reconsider 
this issue in a new perspective based 
on the dispersionless Hirota equations.  
As demonstrated in the case of the dispersionless 
KP and Toda hierarchies \cite{TTZ06}, 
the dispersionless Hirota equations can be 
a powerful tool for studying finite variable reductions 
of dispersionless integrable systems.  
The goal of this paper is to generalize 
this observation to the case of 
the universal Whitham hierarchy.  

This paper is organized as follows.  
Section 2 is a brief review of the universal 
Whitham hierarchy of genus zero.  
Sections 3 and 4 deal with one-variable reduction.  
In Section 3, we derive L\"owner-type equations 
and diagonal hydrodynamic equations as 
necessary conditions for one-variable reduction.  
In Section 4, we show that the hydrodynamic equations 
can be solved by the hodograph method, and lastly 
confirm that the L\"owner-type equations and 
the hydrodynamic equations are sufficient 
to determine a solution of the universal Whitham hierarchy.  
Section 5, 6 and 7 are devoted to multi-variable reductions. 
In Section 5, we examine ``rational reduction'' 
(or ``algebraic orbits'') of the universal Whitham hierarchy 
as a prototype of multi-variable reductions.  
Following the lines illustrated therein, we formulate 
a general form of multi-variable reduction in Section 6, 
and reconsider the results of Guil et al. \cite{GMMA03} 
on the basis of this formulation.  
Section 8 is a summary of our results.

\section{Universal Whitham hierarchy of genus zero}

This section is a collection of basic notions 
and results on the universal Whitham hierarchy 
of genus zero picked out from the literature 
\cite{Krichever94,MMMA0509,MMMA0510,TT06}.

\subsection{Lax equations}

We consider the universal Whitham hierarchy of 
genus zero with $N+1$ marked points on a Riemann sphere 
with coordinate $p$ \cite{Krichever94}.  
One of the marked points are fixed to $p = \infty$; 
the others $p = q_1,\ldots,q_N$ are part of 
the dynamical variables of the hierarchy.  
The other dynamical variables are 
the coefficients of the Laurent series 
\beq
\begin{aligned}
  z_0(p) &= p + \sum_{j=2}^\infty u_{0j}p^{-j+1}, \\
  z_\alpha(p) &= \frac{r_\alpha}{p - q_\alpha} 
    + \sum_{j=1}^\infty u_{\alpha j}(p - q_\alpha)^{j-1}  
    \quad (\alpha = 1,\ldots,N),  
\end{aligned}
\eeq
which are assumed to converge in domains 
$\mathcal{D}_0$ and $\mathcal{D}_\alpha$ 
(or asymptotic expansion of holomorphic functions 
at the boundary points $p = \infty$ and $p = q_\alpha$ 
of such domains).  These Laurent series are 
dispersionless analogues of the Lax operators 
in dispersive integrable hierarchies.  

In this setup, the universal Whitham hierarchy 
has $N+1$ series of time variables 
$t_{\alpha n}$ ($n = 1,2,\ldots$ for $\alpha = 0$ 
and $n = 0,1,\ldots$ for $\alpha = 1,\ldots,N$). 
Let us use $\bst$ to denote these time variables 
collectively, namely, 
\beqnn
  \bst = (t_{01},t_{02},\ldots,t_{10},t_{11},\ldots,
          t_{N0},t_{N1},\ldots). 
\eeqnn
The lowest variables $t_{01},t_{10},\ldots,t_{N0}$ 
in each series $t_{\alpha n}$, $\alpha = 0,1,\ldots,N$, 
play a special role.  In particular, $t_{01}$ should be 
interpreted as a {\it spatial} variable rather than 
a time variable.  In a sense, 
$t_{10},\ldots,t_{N0}$, too, are a kind of 
spatial variables, which originate in 
{\it charge} variables of an $N+1$-component 
charged fermion system \cite{TT06}.  
For convenience, we introduce the auxiliary variable 
\beqnn
  t_{00} = - \sum_{\alpha=1}^N t_{\alpha 0}. 
\eeqnn
(This stems from the constraint of total charge 
being zero.)  We also use the abbreviation 
\beqnn
  \rd_{\alpha n} = \rd/\rd t_{\alpha n}. 
\eeqnn
for the derivatives in the time variables.  

Time evolution of the system is defined by 
the dispersionless Lax equations \cite{Krichever94} 
\beq
  \rd_{\alpha n}z_\beta(p) 
  = \{\Omega_{\alpha n}(p),\, z_\beta(p)\} 
  \quad (\alpha,\beta = 0,1,\ldots,N)
\label{Lax-z(p)}
\eeq
with respect to the two-dimensional Poisson bracket 
\beq
  \{f,g\} 
  = \frac{\rd f}{\rd p}\frac{\rd g}{\rd t_{01}} 
  - \frac{\rd f}{\rd t_{01}}\frac{\rd g}{\rd p} 
\eeq
on the $(p,t_{01})$ space.  Thus $p$ is a conjugate 
variable of $t_{01}$, in other words, 
classical limit of the ``momentum'' $\partial_{01}$ 
in a one-dimensional quantum mechanical system. 
$\Omega_{0n}(p)$'s and $\Omega_{\alpha n}(p)$'s for $n \ge 1$ 
are polynomial in $p$  and $(p - q_\alpha)^{-1}$, 
\beqnn
\begin{aligned}
\Omega_{0n}(p) 
  &= p^n + a_{0n2}p^{n-2} + \cdots + a_{0nn},\\
\Omega_{\alpha n}(p) 
  &= \frac{a_{\alpha n0}}{(p-q_\alpha)^n} 
     + \frac{a_{\alpha n1}}{(p-q_\alpha)^{n-1}}
     + \cdots 
     + \frac{a_{\alpha nn-1}}{(p-q_\alpha)}, 
\end{aligned}
\eeqnn
and given by the singular part of Laurent expansion 
of $z_0(p)^n$ and $z_\alpha(p)^n$ 
(including the constant term for the former): 
\beq
\begin{aligned}
  z_0(p)^n &= \Omega_{0n}(p) + O(p^{-1})
  \quad (p \to \infty), \\
  z_\alpha(p)^n &= \Omega_{\alpha n}(p) + O(1) 
  \quad (p \to q_\alpha). 
\end{aligned}
\eeq
The first few of them read 
\beqnn
\begin{aligned}
& \Omega_{01}(p) = p, \quad 
  \Omega_{02}(p) = p^2 + 2u_{02}, 
  \quad \ldots, \\
& \Omega_{\alpha 1}(p) = \frac{r_\alpha}{p-q_\alpha}, \quad 
  \Omega_{\alpha 2}(p) 
  = \frac{r_\alpha^2}{(p-q_\alpha)^2} 
    + \frac{2r_\alpha u_{\alpha 1}}{p-q_\alpha}, 
  \quad \ldots. 
\end{aligned}
\eeqnn
$\Omega_{\alpha 0}(p)$'s are exceptional 
and given by logarithmic functions: 
\beq
  \Omega_{\alpha 0}(p) = - \log(p - q_\alpha). 
\eeq

\subsection{$S$-functions and Hamilton-Jacobi equations}

We now introduce a set of new variables 
to extend the foregoing Lax equations 
to a larger system.  

The first set of new variables are 
``conjugate variables'' of $z_\beta(p)$'s 
\cite{MMMA0509,MMMA0510}, 
which play the role of the Orlov-Schulman operators 
in the present setup.  
As a consequence of the Lax equations, 
$\Omega_{\alpha n}(p)$'s satisfy 
the dispersionless Zakharov-Shabat equations 
\cite{Krichever94} 
\beq
  \rd_{\beta n}\Omega_{\alpha m}(p) 
  - \rd_{\alpha m}\Omega_{\beta n}(p) 
  + \{\Omega_{\alpha m}(p),\,\Omega_{\beta n}(p)\} 
  = 0  \quad (\alpha,\beta = 0,1,\ldots,N).
\eeq
They can be packed into the equation 
\beq
  \omega \wedge \omega = 0 
\eeq
for the closed 2-form 
\beqnn
  \omega 
  = \sum_{n=1}^\infty d\Omega_{0n}(p)\wedge dt_{0n} 
  + \sum_{\alpha=1}^N\sum_{n=0}^\infty 
      d\Omega_{\alpha n}(p)\wedge dt_{\alpha n}, 
\eeqnn
where ``$d$'' stands for exterior differentiation 
with respect to both $\bst$ and $p$. 
By Darboux's theorem, $\omega$ can be expressed 
in a canonical form with two ``Darboux variables''. 
One can choose $z_\beta(p)$ as one of the Darboux 
variable; let $\zeta_\beta(p)$ denote the conjugate 
variable: 
\beq
  \omega = dz_\beta(p) \wedge d\zeta_\beta(p) 
  \quad (\beta = 0,1,\ldots,N). 
\label{Darboux}
\eeq
This equation implies that $\zeta_\beta(p)$'s 
satisfy Lax equations of the same form 
\beq
  \rd_{\alpha n}\zeta_\beta(p) 
  = \{\Omega_{\alpha n}(p),\, \zeta_\beta(p)\} 
  \quad (\alpha,\beta = 0,1,\ldots,N) 
\eeq
as $z_\beta(p)$'s, along with 
the canonical Poisson relations 
\beq
  \{z_\beta(p),\zeta_\beta(p)\} = 1 
  \quad (\beta = 0,1,\ldots,N). 
\eeq

Having introduced $\zeta_\beta(p)$'s, 
we can now introduce a second set of new variables 
$\calS_\beta(p)$, $\beta = 0,1,\ldots,N$, 
which are called ``$S$-functions'' 
\cite{Krichever94,GMMA03,MMMA0509,MMMA0510}.  
To this end, we rewrite (\ref{Darboux}) as 
\beqnn
  d(\theta + \zeta_\beta(p)dz_\beta(p)) = 0, 
\eeqnn
where 
\beqnn
  \theta = \sum_{n=1}^\infty \Omega_{0n}(p)dt_{0n} 
    + \sum_{\alpha=1}^N\sum_{n=0}^\infty 
      \Omega_{\alpha n}dt_{\alpha n}. 
\eeqnn
The $S$-function $\calS_\beta(p)$ is defined 
as a potential of $\theta + \zeta_\beta(p)dz_\beta(p)$: 
\beq
  \theta + \zeta_\beta(p)dz_\beta(p) = d\calS_\beta(p). 
\eeq
Though one can find these $S$-functions as Laurent series 
of $p$ (for $\beta = 0$) or of $(p - q_\alpha)^{-1}$ 
(for $\beta = 1,\ldots,N$), it is more convenient 
to consider them as Laurent series in $z_\beta(p)$.   
In such an expression, they can be expanded as 
\beq
\begin{aligned}
  \calS_0(p) &= \sum_{n=1}^\infty t_{0n}z_0(p)^n 
    + t_{00}\log z_0(p) 
    - \sum_{n=1}^\infty \frac{z_0(p)^{-n}}{n}v_{0n}, \\
  \calS_\beta(p) &= \sum_{n=1}^\infty t_{\beta n}z_\beta(p)^n 
    + t_{\beta 0}\log z_\beta(p) + \phi_\beta 
    - \sum_{n=1}^\infty \frac{z_\beta(p)^{-n}}{n}v_{\beta n}, 
\end{aligned}
\eeq
where the coefficients $v_{0n},v_{\beta n}$ 
and $\phi_\beta$ are functions of $\bst$. 
We can rewrite this expression as 
\beqnn
  \calS_0(p) = S_0(z_0(p)), \quad 
  \calS_\beta(p) = S_\beta(z_\beta(p)) 
\eeqnn
by introducing the functions 
\beq
\begin{aligned}
  S_0(z) &= \sum_{n=1}^\infty t_{0n}z^n 
    + t_{00}\log z 
    - \sum_{n=1}^\infty \frac{z^{-n}}{n}v_{0n}, \\
  S_\beta(z) &= \sum_{n=1}^\infty t_{\beta n}z^n 
    + t_{\beta 0}\log z + \phi_\beta 
    - \sum_{n=1}^\infty \frac{z^{-n}}{n}v_{\beta n} 
\end{aligned}
\eeq
of $z$.  These $S_\beta(z)$'s, too, are called 
``$S$-functions''.  

In a sense, the second set of $S$-functions $S_\beta(z)$ 
are more fundamental.  They satisfy 
the Hamilton-Jacobi equations \cite{Krichever94,GMMA03} 
\beq
  \rd_{\alpha n}S_\beta(z) 
   = \Omega_{\alpha n}(\rd_{01}S_\beta(z)) 
  \quad (\alpha = 0,1,\ldots,N), 
\label{Hamilton-Jacobi}
\eeq
which are quasiclassical limit of the (scalar-valued) 
auxiliary linear equations of the underlying 
multi-component KP hierarchy \cite{TT06}.  
Moreover, the $t_{01}$-derivative 
\beqnn
  p_\beta(z) = \rd_{01}S_\beta(z) 
\eeqnn
gives the inverse function of $z = z_\beta(p)$, namely, 
\beq
  z_\beta(p_\beta(z)) = z, \quad 
  p_\beta(z_\beta(p)) = p. 
\eeq

We can derive a simple relation among 
$q_\beta,r_\beta$ and $\phi_\beta,v_{\beta 1}$ from 
these remarks.  The foregoing Laurent expansion of 
the $S$-functions implies that $p_0(z)$ and 
$p_\beta(z)$ ($\beta = 1,\ldots,N$) behave as 
\beqnn
  p_0(z) = z + O(z^{-1}), \quad 
  p_\beta(z) 
  = \rd_{01}\phi_0 + \rd_{01}v_{\beta 1}z^{-1} + O(z^{-2}) 
\eeqnn
as $z \to \infty$.  Since the inverse function 
$p = p_\beta(z)$ of 
\beqnn
  z = z_\beta(p) = \frac{r_\beta}{p - q_\beta} + O(1) 
\eeqnn
should coincide with this expression of 
$p_\beta(z)$, we find that 
\beq
  q_\beta = - \rd_{01}\phi_\beta, \quad 
  r_\beta = - \rd_{01}v_{\beta 1}. 
\label{qr-phiv}
\eeq
On the other hand, the Hamilton-Jacobi equations 
for $t_{\alpha 0}$ take such a form as 
\beqnn
  \rd_{\alpha 0}S_\beta(z) 
  = - \log(\rd_{01}S_\beta(z) - q_\alpha) 
  = - \log(p_\beta(z) - q_\alpha), 
\eeqnn
which can be solved for $p_\beta(z)$ as 
\beqnn
  p_\beta(z) = q_\alpha + e^{-\rd_{\alpha 0}S_\beta(z)}. 
\eeqnn
Letting $\alpha = \beta$ and recalling 
the Laurent expansion of $S_\beta(z)$, 
we find that 
\beqnn
  p_\beta(z) 
  = q_\beta + e^{-\rd_{\alpha 0}\phi_0}z^{-1} 
    + O(z^{-2}). 
\eeqnn
This gives another expression of $r_\beta$: 
\beq
  r_\beta = e^{-\rd_{\alpha 0}\phi_\alpha}. 
\eeq

\subsection{$F$-function and Hirota equations}

We now introduce the $F$-function 
$\mathcal{F} = \mathcal{F}(\bst)$ 
(logarithm of the quasiclassical $\tau$ function 
\cite{Krichever94}) as a solution of the equations 
\beq
\begin{aligned}
&\rd_{0n}\mathcal{F} = v_{0n}, \quad 
  \rd_{\alpha n}\mathcal{F} = v_{\alpha n}, \\
&\rd_{\alpha 0}\mathcal{F} 
    = - \phi_\alpha 
      + \sum_{\beta=1}^\alpha t_{\beta 0}\log(-1), 
  \quad \alpha = 1,\ldots,N, 
\end{aligned}
\eeq
where $\log(-1)$ is understood to be equal 
to, say, $\pi i$, though the choice of 
the branch is irrelevant in the final result.  
(This definition of the $F$-function \cite{TT06}
is slightly different from that of 
Ma\~{n}as et al. \cite{MMMA0509,MMMA0510}, 
but this does not affect the main part 
of results.) This strange factor is related 
to the signature factors $\epsilon_{\alpha\beta}$ 
that we shall encounter below.  

All relevant quantities of the hierarchy 
can be expressed in terms of the $F$-function. 
In particular, the $S$-functions have 
a very compact expression: 
\beq
\begin{aligned}
  S_0(z) 
  &= \sum_{n=1}^\infty t_{0n}z^n 
     + t_{00}\log z -  D_0(z)\mathcal{F}, \\
  S_\alpha(z) 
  &= \sum_{n=1}^\infty t_{\alpha n}z^n 
     + t_{\alpha 0}\log z + \phi_\alpha 
     - D_\alpha(z)\mathcal{F}, 
\end{aligned}
\label{S(z)-F}
\eeq
where 
\beqnn
  D_0(z) 
  = \sum_{n=1}^\infty \frac{z^{-n}}{n}\rd_{0n}, \quad 
  D_\alpha(z) 
  = \sum_{n=1}^\infty \frac{z^{-n}}{n}\rd_{\alpha n}. 
\eeqnn
Consequently, the $p$-functions, too, 
can be neatly expressed as 
\beq
\begin{aligned}
  p_0(z) &= z - \rd_{01}D_0(z)\mathcal{F}, \\
  p_\alpha(z) 
  &= - \rd_{01}(D_\alpha(z)+\rd_{\alpha 0})\mathcal{F}. 
\end{aligned}
\label{p(z)-F}
\eeq
The $z$-functions do not have such 
a simple expression, but one can determine 
the Laurent coefficients $u_{\alpha n}$ 
order by order.  For example, let us derive 
an expression of $u_{02},q_\alpha,r_\alpha$.  
The first few terms of the foregoing expression 
of the $p$-functions read 
\beqnn
  p_0(z) &=& z - \rd_{01}^2\mathcal{F} z^{-1} + O(z^{-2}), \\
  p_\alpha(z) &=& - \rd_{01}\rd_{\alpha 0}\mathcal{F} 
    - \rd_{01}\rd_{\alpha 1}\mathcal{F}z^{-1} + O(z^{-2}). 
\eeqnn
On the other hand, as the inverse function of 
\beqnn
  z_0(p) &=& p + u_{02}p^{-1} + O(p^{-2}), \\
  z_\alpha(p) &=& \frac{r_\alpha}{p-q_\alpha} + O(1), 
\eeqnn
$p_0(z)$ and $p_\alpha(z)$ should have 
Laurent expansion of the form 
\beqnn
  p_0(z) &=& z - u_{02}z^{-1} + O(z^{-2}), \\
  p_\alpha(z) &=& q_\alpha + r_\alpha z^{-1} + O(z^{-2}). 
\eeqnn
Thus we find that 
\beq
  u_{02} = \rd_{01}^2\mathcal{F}, \quad 
  q_\alpha = - \rd_{01}\rd_{\alpha 0}\mathcal{F}, \quad 
  r_\alpha = - \rd_{01}\rd_{\alpha 1}\mathcal{F}. 
\label{u02-qr-F}
\eeq
Note that this expression of $q_\alpha$ 
and $r_\alpha$ can also be derived from (\ref{qr-phiv}).  

When the universal Whitham hierarchy is viewed 
as dispersionless (or quasiclassical) limit of 
the (charged) multi-component KP hierarchy \cite{TT06}, 
the $F$-function is identified with the limit 
of logarithm of the tau function.  
One can thereby derive 
the ``dispersionless Hirota equations'' 
(or, more appropriately, ``dispersionless 
differential Fay identities'') 
\beq
\begin{aligned}
e^{\hat{D}_0(z)\hat{D}_0(w)\mathcal{F}}
  &= 1 - \frac{\rd_{01}(\hat{D}_0(z) - \hat{D}_0(w))\mathcal{F}}{z-w},\\
ze^{\hat{D}_0(z)\hat{D}_\alpha(w)\mathcal{F}}
  &= z - \rd_{01}(\hat{D}_0(z) - \hat{D}_\alpha(w))\mathcal{F},\\
e^{\hat{D}_\alpha(z)\hat{D}_\alpha(w)\mathcal{F}} 
  &= - \frac{zw\rd_{01}(\hat{D}_\alpha(z) - \hat{D}_\alpha(w))\mathcal{F}}{z-w},\\
\epsilon_{\alpha\beta}
e^{\hat{D}_\alpha(z)\hat{D}_\beta(w)\mathcal{F}} 
  &= - \rd_{01}(\hat{D}_\alpha(z) - \hat{D}_\beta(w))\mathcal{F} 
     \quad (\alpha \not= \beta)   \\
\end{aligned}
\label{dHirota}
\eeq
for $\alpha,\beta = 1,\ldots,N$, where 
\beqnn
  \hat{D}_\alpha(z) 
= \begin{cases}
  D_0(z) & (\alpha = 0), \\
  D_\alpha(z) + \rd_{\alpha 0} & (\alpha \not= 0), 
  \end{cases}
\qquad 
  \epsilon_{\alpha\beta} 
= \begin{cases}
  +1 & (\alpha \le \beta),\\
  -1 & (\alpha > \beta).  
  \end{cases}
\eeqnn
The signature factor $\epsilon_{\alpha\beta}$ 
stems from a fermionic representation of 
the tau functions of the multi-component KP hierarchy 
\cite{DJKM-III,KvdL-mcKP}. 

These dispersionless Hirota equations are equivalent 
to the universal Whitham hierarchy itself \cite{TT06}. 
This is a generalization of the results known 
for the dispersionless KP and Toda hierarchies 
\cite{TT95,BMRWZ01,Teo03,Teo06}.  
The notion of Faber polynomials and 
Grunsky coefficients, which originate 
in complex analysis, play a crucial role here.

\subsection{Faber polynomials and Grunsky coefficients}

Faber polynomials and Grunsky coefficients 
are hidden in the foregoing setup of 
the universal Whitham hierarchy. 

First of all, $\Omega_{0n}(z)$ and 
$\Omega_{\alpha n}(z)$ are nothing but 
the Faber polynomials of $p_0(z)$ and $p_\alpha(z)$, 
respectively \cite{TT06}.  Namely, these ``polynomials'' 
(in $p$ and in $(p - q_\alpha)^{-1}$, respectively) 
are characterized by the following generating functions: 
\beq
\begin{aligned}
\log\frac{p_0(z) - q}{z} 
  &= - \sum_{n=1}^\infty 
       \frac{z^{-n}}{n}\Omega_{0n}(q), \\
\log\frac{q - p_\alpha(z)}{q - q_\alpha} 
  &= - \sum_{n=1}^\infty 
       \frac{z^{-n}}{n}\Omega_{\alpha n}(q). 
\end{aligned}
\label{Faber}
\eeq
Moreover, differentiating these generating functions 
by $q$ yields the generating functions 
\beq
\begin{aligned}
\frac{1}{q - p_0(z)} 
  &= - \sum_{n=1}^\infty 
       \frac{z^{-n}}{n}\Omega_{0n}'(q), \\
\frac{1}{q-p_\alpha(z)} - \frac{1}{q-q_\alpha} 
  &= - \sum_{n=1}^\infty
       \frac{z^{-n}}{n}\Omega_{\alpha n}'(q) 
\end{aligned}
\label{kernel}
\eeq
for the derivatives 
\beqnn
  \Omega_{0n}'(q) = \frac{\rd\Omega_{0n}(q)}{\rd q}, 
  \quad 
  \Omega_{\alpha n}'(q) 
  = \frac{\rd\Omega_{\alpha n}(q)}{\rd q}. 
\eeqnn
It will be more suggestive (and convenient) 
to rewrite the second equation of (\ref{kernel}) as 
\beq
  \frac{1}{q-p_\alpha(z)} 
  = - \sum_{n=1}^\infty 
      \frac{z^{-n}}{n}\Omega_{\alpha 0}'(q) 
    - \Omega_{\alpha 0}'(q). 
\label{kernel2}
\eeq
This type of identities are called ``kernel formulas'' 
in the literature \cite{CK95,CT06}
(because the left hand side 
may be thought of as a Cauchy kernel), 
and shown to be useful in some applications 
of dispersionless Hirota equations such as 
``associativity equations'' \cite{BMRWZ01,CT06}.  

Grunsky coefficients show up  when we use (\ref{p(z)-F}) 
to rewrite the right hand side of 
the dispersionless Hirota equations (\ref{dHirota}) as 
\beqnn
\begin{aligned}
1 - \frac{\rd_{01}(\hat{D}_0(z) - \hat{D}_0(w))\mathcal{F}}{z-w}
  &= p_0(z) - p_0(w), \\
z - \rd_{01}(\hat{D}_0(z) - \hat{D}_\alpha(w))\mathcal{F} 
  &= p_0(z) - p_\alpha(w), \\
- \frac{zw\rd_{01}(\hat{D}_\alpha(z) - \hat{D}_\alpha(w))\mathcal{F}}{z-w}
  &= \frac{zw(p_\alpha(z) - p_\alpha(w))}{z-w}, \\
- \rd_{01}(\hat{D}_\alpha(z) - \hat{D}_\beta(w))\mathcal{F} 
  &= p_\alpha(z) - p_\beta(w), 
\end{aligned}
\eeqnn
and consider the logarithm of both hand sides 
of each equation.  (\ref{dHirota}) thus turn 
into the following equations 
\beq
\begin{aligned}
\hat{D}_0(z)\hat{D}_0(w)\mathcal{F} 
  &= \log\frac{p_0(z)-p_0(w)}{z-w},\\
\hat{D}_0(z)\hat{D}_0(w)\mathcal{F} 
  &= \log\frac{p_0(z)-p_\alpha(w)}{z},\\
\hat{D}_\alpha(z)\hat{D}_\alpha(w)\mathcal{F} 
  &= \log\frac{zw(p_\alpha(z)-p_\alpha(w))}{z-w},\\
\hat{D}_\alpha(z)\hat{D}_\beta(w)\mathcal{F} 
  &= \log\frac{p_\alpha(z)-p_\beta(w)}{\epsilon_{\alpha\beta}} 
     \quad (\alpha \not= \beta). 
\end{aligned}
\label{log-dHirota}
\eeq
As one can show using the definition of 
the Faber polynomials (\ref{Faber}), 
these equations are, actually, 
a generating functional representation 
of the Hamilton-Jacobi equations 
(\ref{Hamilton-Jacobi}); this fact lies 
in the heart of the aforementioned equivalence 
of the dispersionless Hirota equations 
and the universal Whitham hierarchy.  

The right hand side of (\ref{log-dHirota}) 
are nothing but generating functions of 
the (generalized) Grunsky coefficients 
$b_{\alpha m\beta n}$ of the $p$-functions: 
\beq
\begin{aligned}
\log\frac{p_0(z)-p_0(w)}{z-w} 
  &= - \sum_{m,n=1}^\infty z^{-m}w^{-n}b_{0m0n},\\
\log\frac{p_0(z)-p_\alpha(w)}{z} 
  &= - \sum_{m=1}^\infty\sum_{n=0}^\infty 
       z^{-m}w^{-n}b_{0m\alpha n},\\
\log\frac{zw(p_\alpha(z)-p_\alpha(w))}{z-w} 
  &= - \sum_{m,n=0}^\infty z^{-m}w^{-n}b_{\alpha m\alpha n},\\
\log\frac{p_\alpha(z)-p_\beta(w)}{\epsilon_{\alpha\beta}} 
  &= - \sum_{m,n=0}^\infty z^{-m}w^{-n}b_{\alpha m\beta n} 
  \quad (\alpha \not= \beta). 
\end{aligned}
\label{Grunsky}
\eeq
In particular, the $z^{-1}$ terms of 
the first and second equations yield the identities 
\beq
  w - p_0(w) = \sum_{n=1}^\infty w^{-n}b_{010n}, \quad 
  - p_\alpha(w) = - \sum_{n=0}^\infty w^{-n}b_{01\alpha n}. 
\eeq
One can thus recover the $p$-functions from 
the Grunsky coefficients.  

(\ref{log-dHirota}) show that all second derivatives 
of the $F$-function are given by the Grunsky coefficients 
as 
\beq
  \hat{\rd}_{\alpha m}\hat{\rd}_{\beta n}\mathcal{F} 
  = - b_{\alpha m\beta n} 
  \quad (\alpha,\beta = 0,1,\ldots,N), 
\label{ddF}
\eeq
where we have introduce the rescaled derivatives 
\beqnn
  \hat{\rd}_{\alpha n} 
= \begin{cases}
  \frac{1}{n}\rd_{\alpha n} & (n \not= 0), \\
  \rd_{\alpha 0} &(n = 0). 
  \end{cases}
\eeqnn
Conversely, one can use (\ref{ddF}) 
as defining equations of $F$.  
In that case, the Grunsky coefficients 
have to satisfy a set of integrability conditions: 
\beq
  \hat{\rd}_{\gamma n}b_{\alpha l\beta m} 
= \hat{\rd}_{\alpha l}b_{\gamma n\beta m}. 
\label{ddF-integrable}
\eeq
Once those integrability conditions are shown 
to be satisfied, one can obtain the $F$-function 
as a solution of these equations.   Moreover, 
this simultaneously ensures that the dispersionless 
Hirota equations are also satisfied, because 
(\ref{ddF}) are, after all, the dispersionless 
Hirota equations in disguise.  
This method for showing the existence of 
the $F$-function was first developed for the case 
of the dispersionless KP and Toda hierarchies 
\cite{TTZ06}.  We shall apply this method 
to reductions of the universal Whitham hierarchy.

\section{L\"owner-type equations in one-variable reduction}

In the setup of one-variable reduction, 
the dynamical variables $u_{\alpha n}$ ($n = 0,1,\ldots$, 
$u_{\alpha 0} = r_\alpha$) and $q_\alpha$ 
are assumed to depend on $\bst$ via 
a single function $\lambda = \lambda(\bst)$ as 
\beqnn
  u_{\alpha n} = u_{\alpha n}(\lambda(\bst)), \quad 
  q_\alpha = q_\alpha(\lambda(\bst)).  
\eeqnn
Consequently, $z_\alpha(p)$ and $p_\alpha(z)$ 
are thought of as functions of $p$ and 
$\lambda(\bst)$: 
\beqnn
  z_\alpha(p) = z_\alpha(p,\lambda(\bst)), \quad 
  p_\alpha(z) = p_\alpha(z,\lambda(\bst)). 
\eeqnn
The goal of this section is to derive the following 

\begin{theorem}\label{th:Loewner}
If $u_{\alpha n}$ and $q_\alpha$ are functions 
of a single variable $\lambda = \lambda(\bst)$, 
then there is a function $U= U(\lambda)$ of $\lambda$ 
such that $p_\alpha(z) = p_\alpha(z,\lambda)$ 
and $z_\alpha(p) = z_\alpha(p,\lambda)$, 
$\alpha = 0,1,\ldots,N$, satisfy 
the L\"owner-type equations 
\beq
  \frac{\rd p_\alpha(z)}{\rd\lambda} 
&=& \frac{1}{U - p_\alpha(z)}
    \frac{\rd u_{02}}{\rd\lambda}, 
\label{Loewner-p(z)}\\
  \frac{\rd z_\alpha(p)}{\rd\lambda}
&=& \frac{z_\alpha'(p)}{p - U}
    \frac{\rd u_{02}}{\rd\lambda}, 
\label{Loewner-z(p)}
\eeq
where $z_\alpha'(p)$ denotes the $p$-derivative 
\beqnn
  z_\alpha'(p) = \frac{\rd z_\alpha(p)}{\rd p}. 
\eeqnn
Moreover, $\lambda = \lambda(\bst)$ satisfies 
the hydrodynamic equations
\beq
  \rd_{\alpha n}\lambda 
  = \chi_{\alpha n}(\lambda)\rd_{01}\lambda 
\label{dlambda}
\eeq
with characteristic speeds 
\beqnn
  \chi_{\alpha n}(\lambda) 
  = \Omega_{\alpha n}'(U). 
\eeqnn
\end{theorem}

This theorem is a generalization of 
the one-variable reduction of the KP and Toda 
hierarchies \cite{TTZ06}.  
Let us explain some implications: 

\begin{itemize}
\item[1.]
In a quite general situation as the theorem assumes, 
we can say nothing about the shape of the domain of 
$p_\alpha(z)$ (equivalently, the range of $z_\alpha(p)$) 
and the position of $U$ therein.  In particular, 
it is {\it a priori} not evident whether the domain 
of $p_\alpha(z)$ is a slit domain like the one 
assumed in the original setup of the L\"owner equation.   
Since the reduction defined by (\ref{Loewner-p(z)}) 
and (\ref{Loewner-z(p)}) seems to be meaningful 
in such a general situation (cf. various solutions 
of L\"owner-type solutions known in the literature
\cite{YG00,BG03,BG04,MMAM02,GMMA03,Manas04,TTZ06}), 
we do not restrict our consideration to slit domains.  
\item[2.] 
All equations of (\ref{Loewner-p(z)}) and 
(\ref{Loewner-z(p)}) have a common driving force $U$.  
In particular, (\ref{Loewner-z(p)}) mean that 
$z_\alpha(p)$'s satisfy an identical 
linear differential equation of the form 
\beqnn
  \left(\frac{\rd}{\rd\lambda} 
    - \frac{1}{p-U}\frac{\rd u_{02}}{\rd\lambda}
      \frac{\rd}{\rd p}\right) z_\alpha(p) = 0.
\eeqnn
By the classical theory of characteristics, 
this implies that $z_\alpha(p)$'s are mutually 
functionally related.  Namely, there are 
functions $f_\alpha(z)$ of one variable $z$ 
such that 
\beq
  z_\alpha(p) = f_\alpha(z_0(p)), 
  \quad \alpha = 1,\ldots,N. 
\label{zalpha(p)-z0(p)}
\eeq
This agrees with the setup of reductions 
by Guil et al. \cite{GMMA03}, 
though they assume the special form 
\beq
  f_\alpha(z) = \frac{1}{z - c_\alpha}, 
\label{GMMA03-assume}
\eeq
where $c_\alpha$'s are constants.  
\item[3.] 
(\ref{dlambda}) determine the time evolution 
of the reduced dynamical variable 
$\lambda = \lambda(\bst)$.  The whole hierarchy 
is thus reduced to the hydrodynamic equations  
(\ref{dlambda}) with Riemann invariant $\lambda$ 
and characteristic speeds  $\chi_{\alpha n}(\lambda)$.  
These hydrodynamic equations can be solved 
by the hodograph method (see Theorem \ref{th:hodograph}).
\item[4.]
It is not obvious from the subsequent proof 
that the converse of the statement of the theorem 
also holds.  Therefore one has to show, separately, 
that (\ref{Loewner-p(z)}) (or (\ref{Loewner-z(p)})) 
and (\ref{dlambda}) lead to a solution of 
the universal Whitham hierarchy.  We shall do it 
with the aid of the dispersionless Hirota equations 
(see Theorem \ref{th:integrability}). 
\end{itemize}

\subsection{Proof of Theorem \ref{th:Loewner}: step 1}

Differentiating both hand sides of the first equation 
of (\ref{log-dHirota}) by $t_{01}$ yields the equation 
\beqnn
 \rd_{01}D_0(z)D_0(w)\mathcal{F} 
  = \frac{\rd_{01}p_0(z)-\rd_{01}p_0(w)}{p_0(z)-p_0(w)}. 
\eeqnn
Note that $\hat{D}_0(z) = D_0(z)$.  
Let us rewrite this equation as 
\beqnn
\begin{aligned}
p_0(z) - p_0(w) 
 &= \frac{\rd_{01}p_0(z)-\rd_{01}p_0(w)}
         {\rd_{01}D_0(z)D_0(w)\mathcal{F}} \\
 &= - \frac{\rd_{01}p_0(z)}{D_0(w)p_0(z)} 
    + \frac{\rd_{01}p_0(w)}{D_0(z)p_0(w)}. 
\end{aligned}
\eeqnn
We have used (\ref{p(z)-F}) to rewrite 
the denominator $\rd_{01}D_0(z)D_0(w)\mathcal{F}$ as 
\beqnn
  \rd_{01}D_0(z)D_0(w)\mathcal{F} 
  = - D_0(w)p_0(z) = - D_0(z)p_0(w). 
\eeqnn
By the chain rule, we can express the derivatives 
on the right hand side as 
\beqnn
\begin{aligned}
\rd_{01}p_0(z) 
  &= \frac{\rd p_0(z)}{\rd\lambda}\rd_{01}\lambda,&
\rd_{01}p_0(w) 
  &= \frac{\rd p_0(w)}{\rd\lambda}\rd_{01}\lambda,\\
D_0(w)p_0(z) 
  &= \frac{\rd p_0(z)}{\rd\lambda}D_0(w)\lambda,&
D_0(z)p_0(w) 
  &= \frac{\rd p_0(w)}{\rd\lambda}D_0(z)\lambda. 
\end{aligned}
\eeqnn
Hence last equation reduces to 
\beqnn
  p_0(z) - p_0(w) 
= - \frac{\rd_{01}\lambda}{D_0(w)\lambda} 
  + \frac{\rd_{01}\lambda}{D_0(z)\lambda} 
\eeqnn
or, equivalently, 
\beqnn
  p_0(z) - \frac{\rd_{01}\lambda}{D_0(z)\lambda} 
= p_0(w) - \frac{\rd_{01}\lambda}{D_0(w)\lambda}. 
\eeqnn
Therefore both hand sides of this equation 
are independent of $z$ and $w$.  
Let $U_0 = U_0(\lambda)$ denote this quantity: 
\beq
  p_0(z) - \frac{\rd_{01}\lambda}{D_0(z)\lambda} 
  = U_0. 
\label{Loewner-pr-1a}
\eeq

On the other hand, applying $D_0(z)$ 
to both hand sides of the first formula 
of (\ref{u02-qr-F}) and using (\ref{p(z)-F})  
yield the identity 
\beqnn
  D_0(z)u_{02} = \rd_{01}^2D_0(z)\mathcal{F} = - \rd_{01}p_0(z). 
\eeqnn
Again by the chain rule, the derivatives 
on both hand sides can be expressed as 
\beqnn
D_0(z)u_{02} 
  = \frac{\rd u_{02}}{\rd\lambda}D_0(z)\lambda, \quad 
\rd_{01}p_0(z) 
  = \frac{\rd p_0(z)}{\rd\lambda}\rd_{01}\lambda 
\eeqnn
Thus we find that 
\beq
  \frac{\rd_{01}\lambda}{D_0(z)\lambda}
  = - \frac{\rd u_{02}/\rd\lambda}{\rd p_0(z)/\rd\lambda}. 
\label{Loewner-pr-1b}
\eeq

We can use (\ref{Loewner-pr-1b}) to rewrite 
the foregoing equation (\ref{Loewner-pr-1a}) as 
\beqnn
  p_0(z) 
  + \frac{\rd u_{02}/\rd\lambda}{\rd p_0(z)/\rd\lambda} 
  = U_0. 
\eeqnn
This implies that $p_0(z)$ satisfies 
the L\"owner-type equation 
\beq
  \frac{\rd p_0(z)}{\rd\lambda}
  = \frac{1}{U_0 - p_0(z)}\frac{\rd u_{02}}{\rd\lambda}. 
\label{Loewner-pr-1c}
\eeq

\subsection{Proof of Theorem \ref{th:Loewner}: step 2}

We can repeat almost the same calculations 
for the second equation of (\ref{log-dHirota}). 

Firstly, differentiating both hand sides 
by $t_{01}$, we obtain the equation 
\beqnn
  \rd_{01}\hat{D}_\alpha(z)\hat{D}_\alpha(w)\mathcal{F} 
  = \frac{\rd_{01}p_\alpha(z)-\rd_{01}p_\alpha(w)}
         {p_\alpha(z)-p_\alpha(w)}, 
\eeqnn
which can be rewritten as 
\beqnn
\begin{aligned}
  p_\alpha(z) - p_\alpha(w) 
&= \frac{\rd_{01}p_\alpha(z)-\rd_{01}p_\alpha(w)}
        {\rd_{01}\hat{D}_\alpha(z)\hat{D}_\alpha(w)\mathcal{F}} \\
&= - \frac{\rd_{01}p_\alpha(z)}{\hat{D}_\alpha(z)p_\alpha(z)} 
   + \frac{\rd_{01}p_\alpha(w)}{\hat{D}_\alpha(w)p_\alpha(w)}. 
\end{aligned}
\eeqnn
By the chain rule, this equation reduces to 
\beqnn
  p_\alpha(z) 
  - \frac{\rd_{01}\lambda}{\hat{D}_\alpha(z)\lambda} 
= p_\alpha(w) 
  - \frac{\rd_{01}\lambda}{\hat{D}_\alpha(w)\lambda}. 
\eeqnn
This implies that both hand sides are 
actually independent of $z$ and $w$.  
Thus we have the equation 
\beq
  p_\alpha(z) 
  - \frac{\rd_{01}\lambda}{\hat{D}_\alpha(z)\lambda}
  = U_\alpha, 
\label{Loewner-pr-2a}
\eeq
where $U_\alpha = U_\alpha(\lambda)$ is a function 
of $\lambda$ only.  

  Secondly, we can derive, from the first formula 
of (\ref{u02-qr-F}), the identity 
\beqnn
  \hat{D}_\alpha(z)u_{02} 
  = \rd_{01}^2\hat{D}_\alpha(z)\mathcal{F} 
  = - \rd_{01}p_\alpha(z). 
\eeqnn
By the chain rule, this identity reduces to 
\beq
  \frac{\rd_{01}\lambda}{\hat{D}_\alpha(z)}
  = - \frac{\rd u_{02}/\rd\lambda}
           {\rd p_\alpha(z)/\rd\lambda}. 
\label{Loewner-pr-2b}
\eeq

(\ref{Loewner-pr-2a}) and (\ref{Loewner-pr-2b}) 
imply that $p_\alpha(z)$ satisfies 
the L\"owner-type equation 
\beq
  \frac{\rd p_\alpha(z)}{\rd\lambda} 
  = \frac{1}{U_\alpha - p_\alpha(z)}\frac{\rd u_{02}}{\rd\lambda}. 
\label{Loewner-pr-2c}
\eeq

\subsection{Proof of Theorem \ref{th:Loewner}: step 3}

Let us examine implications of the third and fourth 
equation of (\ref{log-dHirota}).  

Differentiating the third equation by $t_{01}$ 
yields the equation 
\beqnn
  p_0(z) - p_\alpha(w) 
= - \frac{\rd_{01}p_0(z)}{\hat{D}_\alpha(w)p_0(z)} 
  + \frac{\rd_{01}p_\alpha(w)}{D_0(z)p_\alpha(w)}. 
\eeqnn
By the chain rule, this equation reduces to 
\beqnn
  p_0(z) - p_\alpha(w) 
= - \frac{\rd u_{02}/\rd\lambda}{\rd o_0(z)/\rd\lambda} 
  + \frac{\rd u_{02}/\rd\lambda}{\rd p_\alpha(w)/\rd\lambda}. 
\eeqnn
By (\ref{Loewner-pr-1c}) and (\ref{Loewner-pr-2c}), 
we can rewrite the two terms 
on the right hand side as 
\beqnn
  \frac{\rd u_{02}/\rd\lambda}{\rd p_0(z)/\rd\lambda} 
  = U_0 - p_0(z), \quad 
  \frac{\rd u_{02}/\rd\lambda}{\rd p_\alpha(w)/\rd\lambda}
  = U_\alpha - p_\alpha(w). 
\eeqnn
Thus we find that 
\beq
  U_0 = U_\alpha, \quad \alpha = 1,\ldots,N. 
\eeq
namely, (\ref{Loewner-pr-1c}) and (\ref{Loewner-pr-2c}) 
actually have an identical driving force $U$.  

The same conclusion follows from the fourth equation 
of (\ref{log-dHirota}). Namely, 
differentiating this equation by $t_{01}$ 
eventually leads to the identities $U_\alpha = U_\beta$.

\subsection{Proof of Theorem \ref{th:Loewner}: step4}

We can derive the evolution equations (\ref{dlambda}) 
as follows.  

Let us rewrite (\ref{Loewner-pr-1b}) as 
\beqnn
  D_0(z)\lambda 
  = - \frac{\rd p_0(z)/\rd\lambda}
      {\rd u_{02}/\rd\lambda}\rd_{01}\lambda 
  = - \frac{\rd_{01}\lambda}{U - p_0(z)} 
\eeqnn
and use the identity 
\beqnn
  \frac{1}{U - p_0(z)} 
  = - \sum_{n=1}^\infty \frac{z^{-n}}{n}\Omega_{0n}'(U)
\eeqnn
that can be obtained from the kernel formula 
(\ref{kernel}) by letting $q = U$.  
The outcome is the equation
\beqnn
  D_0(z)\lambda 
  = \sum_{n=1}^\infty 
    \frac{z^{-n}}{n}\Omega_{0n}'(U)\rd_{01}\lambda. 
\eeqnn
This is a generating functional form 
of the equations of (\ref{dlambda}) 
for $\alpha = 0$, $n = 1,2,\ldots$.  

In exactly the same way, using (\ref{Loewner-pr-2b}) 
and the kernel formula for $p_\alpha(z)$, 
we can derive the equations 
\beqnn
  \hat{D}_\alpha(z)\lambda 
  = (D_\alpha(z)+\rd_{\alpha 0})\lambda 
  = \left(\sum_{n=1}^\infty 
      \frac{z^{-n}}{n}\Omega_{\alpha n}'(U) 
      + \Omega_{\alpha 0}'(U)\right)\rd_{01}\lambda. 
\eeqnn
They are a generating functional form 
of the equations of (\ref{dlambda}) 
for $\alpha = 1,\ldots,N$ and $n = 0,1,\ldots$.

\section{Solutions from one-variable reduction}

\subsection{Hodograph method} 

The hydrodynamic equations (\ref{dlambda}) 
can be solved by the hodograph method. 
The hodograph method is extremely simplified 
in this case, because there is only one 
variable $\lambda$.  

\begin{theorem}\label{th:hodograph}
Let $F(\lambda)$ be an arbitrary function 
of $\lambda$, and $\lambda = \lambda(\bst)$ 
a function that satisfies the hodograph equation 
\beq
  \sum_{n=1}^\infty t_{0n}\chi_{0n}(\lambda) 
  + \sum_{\alpha=1}^N\sum_{n=0}^\infty 
    t_{\alpha n}\chi_{\alpha n}(\lambda) 
  = F(\lambda). 
\eeq
Further assume that the regularity condition 
\beq
  \sum_{n=1}^\infty 
    t_{0n}\frac{\rd\chi_{0n}(\lambda)}{\rd\lambda} 
  + \sum_{\alpha=1}^N\sum_{n=0}^\infty 
    t_{\alpha n}\frac{\rd\chi_{\alpha n}(\lambda)}{\rd\lambda}
  \not= \frac{\rd F(\lambda)}{\rd\lambda}  
\eeq
holds for $\lambda = \lambda(\bst)$.  
Then $\lambda = \lambda(\bst)$ satisfies 
the hydrodynamic equations (\ref{dlambda}). 
\end{theorem}

\paragraph*{Proof} 
We differentiating both hand sides of 
the hodograph equation by $t_{\alpha n}$. 
By the chain rule, this yields the equations
\beqnn
  \chi_{\alpha n}(\lambda) 
  + \left(
      \sum_{m=1}^\infty 
      t_{0m}\frac{\rd\chi_{0m}(\lambda)}{\rd\lambda}
    + \sum_{\beta=1}^N\sum_{m=0}^\infty 
      t_{\beta m}\frac{\rd\chi_{\beta m}(\lambda)}{\rd\lambda}
    \right)\rd_{\alpha n}\lambda 
  = \frac{\rd F(\lambda)}{\rd\lambda}\rd_{\alpha n}\lambda, 
\eeqnn
hence 
\beqnn
  \chi_{\alpha n}(\lambda) 
  = \left(
      \frac{\rd F(\lambda)}{\rd\lambda}
    - \sum_{m=1}^\infty 
      t_{0m}\frac{\rd\chi_{0m}(\lambda)}{\rd\lambda}
    - \sum_{\beta=1}^N\sum_{m=0}^\infty 
      t_{\beta m}\frac{\rd\chi_{\beta m}(\lambda)}{\rd\lambda}
    \right)\rd_{\alpha n}\lambda. 
\eeqnn
In particular, letting $\alpha = 0$ and $n =0$, 
we have the equation 
\beqnn
1 = \left(
    \frac{\rd F(\lambda)}{\rd\lambda}
    - \sum_{m=1}^\infty 
      t_{0m}\frac{\rd\chi_{0m}(\lambda)}{\rd\lambda}
    - \sum_{\beta=1}^N\sum_{m=0}^\infty 
      t_{\beta m}\frac{\rd\chi_{\beta m}(\lambda)}{\rd\lambda}
    \right)\rd_{01}\lambda.  
\eeqnn
If we multiply the last equation by 
$\chi_{\alpha n}(\lambda)$ and subtract 
it from the previous one, the outcome 
is the equation 
\beqnn
\begin{aligned}
0 &= \left(
    \frac{\rd F(\lambda)}{\rd\lambda}
    - \sum_{m=1}^\infty 
      t_{0m}\frac{\rd\chi_{0m}(\lambda)}{\rd\lambda}
    - \sum_{\beta=1}^N\sum_{m=0}^\infty 
      t_{\beta m}\frac{\rd\chi_{\beta m}(\lambda)}{\rd\lambda}
    \right) \times {}\\
  &\quad \times 
    (\rd_{\alpha n}\lambda - \chi_{\alpha n}(\lambda)\rd_{01}\lambda). 
\end{aligned}
\eeqnn
By the regularity condition, we can drop the prefactor 
of $\rd_{\alpha n}\lambda - \chi_{\alpha n}(\lambda)\rd_{01}\lambda$ 
and obtain the hydrodynamic equations (\ref{dlambda}). 
\qed

\subsection{Existence of $F$-function}

In the last section, we derived 
the L\"owner-type equations (\ref{Loewner-p(z)}), 
(\ref{Loewner-z(p)}) and the hydrodynamic equations 
(\ref{dlambda}), but we have not confirmed the converse, 
namely, whether these equations ensure 
that $u_{\alpha n} = u_\alpha(\lambda(\bst))$ 
and $q_\alpha = q_\alpha(\lambda(\bst))$ give 
a solution of the universal Whitham hierarchy. 

We now prove that the converse is also true, 
following the idea presented in the end of 
Section 2.4. This is also a generalization of 
the result  in the case of the dispersionless 
KP and Toda hierarchies \cite{TTZ06}. 

\begin{theorem}\label{th:integrability}
The integrability conditions (\ref{ddF-integrable}) 
of (\ref{ddF}) are satisfied in the foregoing setup 
of one-variable reduction.  The $F$-function 
$\mathcal{F} = \mathcal{F}(\bst)$ thus defined 
by (\ref{ddF}) gives a solution of 
the dispersionless Hirota equations (\ref{dHirota}). 
\end{theorem}

\paragraph*{Proof}
Let us first illustrate the calculations 
in the case where $\alpha = \beta = \gamma = 0$.  
We substitute $z =z_1$ and $w = z_2$ 
in the first generating function of (\ref{Grunsky}), 
and apply $\hat{D}_0(z_3) = D_0(z_3)$ to both hand sides.  
This yields the generating function 
\beqnn
  \frac{D_0(z_3)(p_0(z_1)-p_0(z_2))}{p_0(z_1)-p_0(z_2)}
  = - \sum_{l,m,n=1}^\infty 
      z_1^{-l}z_2^{-m}z_3^{-n}\hat{\rd}_{0n}b_{0l0m}
\eeqnn
of $\hat{\rd}_{0n}b_{0l0m}$'s.  
On the other hand, recall that 
the hydrodynamic equations (\ref{dlambda}) 
have a generating functional representation 
(cf. step 4 of the proof of Theorem \ref{th:Loewner}). 
In the case of $\alpha = 0$, it reads 
\beqnn
  D_0(z)\lambda = \frac{\rd_{01}\lambda}{p_0(z) - U}. 
\eeqnn
Using this equation, the L\"owner-type equation 
for $p_0(z)$ and the chain rule, 
we can rewrite the left hand side of 
the foregoing generating function of 
$\hat{\rd}_{0n}b_{0l0m}$'s as 
\beqnn
\begin{aligned}
&\frac{D_0(z_3)(p_0(z_1)-p_0(z_2))}{p_0(z_1)-p_0(z_2)} \\
&= \frac{D_0(z_3)\lambda}{p_0(z_1)-p_0(z_2)} 
   \left(\frac{\rd p_0(z_1)}{\rd\lambda} 
       - \frac{\rd p_0(z_2)}{\rd\lambda}\right) \\
&= \frac{\rd_{01}\lambda}
   {(p_0(z_3)-U)(p_0(z_1)-p_0(z_2))} 
   \left(\frac{1}{U-p_0(z_1)}-\frac{1}{U-p_0(z_2)}\right) 
   \frac{\rd u_{02}}{\rd\lambda} \\
&= - \frac{\rd_{01}\lambda}{(U-p_0(z_1))(U-p_0(z_2))(U-p_0(z_3))} 
     \frac{\rd u_{02}}{\rd\lambda} .
\end{aligned}
\eeqnn
Since this quantity is symmetric in $z_1$ and $z_3$, 
we have the functional identity 
\beqnn
  \frac{D_0(z_3)(p_0(z_1)-p_0(z_2))}{p_0(z_1)-p_0(z_2)} 
= \frac{D_0(z_1)(p_0(z_3)-p_0(z_2))}{p_0(z_3)-p_0(z_2)}. 
\eeqnn
This implies the identities 
\beqnn
  \hat{\rd}_{0n}b_{0l0m} = \hat{\rd}_{0l}b_{0n0m} 
\eeqnn
of the coefficients, which are exactly 
the integrability conditions (\ref{ddF-integrable}) 
for $\alpha = \beta = \gamma = 0$. 

In much the same way, staring with one of 
the generating functions of (\ref{Grunsky}) 
and applying $\hat{D}_\gamma(z)$ to it, 
we obtain the generating functions 
\beqnn
  \frac{\hat{D}_\gamma(z_3)(p_\alpha(z_1)-p_\beta(z_2))}
       {p_\alpha(z_1)-p_\beta(z_2)} 
= - \sum_{l,m,n}z_1^{-l}z_2^{-m}z_3^{-n} 
    \hat{\rd}_{\gamma n}b_{\alpha l\beta m} 
\eeqnn
of derivatives of the general Grunsky coefficients. 
Since the hydrodynamic equations (\ref{dlambda}) 
have the generating functional form 
\beqnn
  \hat{D}_\alpha(z)\lambda 
  = \frac{\rd_{01}\lambda}{p_\alpha(z)-U}, 
\eeqnn
we can rewrite the foregoing generating functions as 
\beqnn
  \frac{\hat{D}_\gamma(z_3)(p_\alpha(z_1)-p_\beta(z_2))}
       {p_\alpha(z_1)-p_\beta(z_2)} 
= - \frac{\rd_{01}\lambda}
    {(U-p_\alpha(z_1))(U-p_\beta(z_2))(U-p_\gamma(z_3))}
    \frac{\rd u_{02}}{\rd\lambda}. 
\eeqnn
This implies the functional identity 
\beqnn
  \frac{\hat{D}_\gamma(z_3)(p_\alpha(z_1)-p_\beta(z_2))}
       {p_\alpha(z_1)-p_\beta(z_2)} 
= \frac{\hat{D}_\alpha(z_1)(p_\gamma(z_3)-p_\beta(z_2))}
       {p_\gamma(z_3)-p_\beta(z_2)}, 
\eeqnn
hence the integrability conditions  
(\ref{ddF-integrable}) as expected.  
More precisely, we have to be careful 
about the difference of the four types 
of generating functions in (\ref{Grunsky}),  
but this does not affect the final conclusion.  
\qed

Let us mention that this proof reveals 
an interesting feature of the integrability 
conditions (\ref{ddF-integrable}).  Namely, 
since the derivatives of the Grunsky coefficients 
are nothing but the third derivatives of 
the $F$-function, the functional identities 
in the proof imply that these third derivatives 
have a generating function of the form 
\beq
\begin{aligned}
& \sum_{l,m,n}z_1^{-l}z_2^{-m}z_3^{-n}
  \hat{\rd}_{\alpha l}\hat{\rd}_{\beta m}\hat{\rd}_{\gamma n}\mathcal{F}\\
&= - \frac{\rd_{01}\lambda}
     {(U-p_\alpha(z_1))(U-p_\beta(z_2))(U-p_\gamma(z_3))}
     \frac{\rd u_{02}}{\rd\lambda}. 
\end{aligned}
\label{dddF}
\eeq
This result is suggestive from the point of view of 
associativity equations \cite{BMRWZ01,CT06}, 
because the third derivatives are 
fundamental quantities therein.  We shall see 
that a similar result holds in multi-variable reductions 
as well.

\section{Rational reductions (algebraic orbits)}

Unfortunately, it seems difficult to extend 
the foregoing method for the one-variable reduction 
to multi-variable reductions.  
In the case of multi-variable reductions, 
we shall {\it start} from L\"owner-type equations 
rather than {\it derive} them.  
The problem is how to find a correct form of 
L\"owner-type equations.   
Though an answer to this question is presented 
in the work of Guil et al. \cite{GMMA03}, 
we dare to take a different (heuristic) route 
that leads to the same answer.  

Our strategy is to examine rational reductions 
of the universal Whitham hierarchy  
(or ``algebraic orbits'' in the terminology 
of Krichever \cite{Krichever94}) as a prototype 
of general multi-variable reductions.  
This class of reductions cover, 
for example, the Zakharov reduction of 
the Benney equations \cite{Zakharov80}, 
reductions of the dispersionless KP hierarchy 
related to 2D topological field theories 
\cite{Krichever91,Dubrovin92,AK96}, 
a hydrodynamic reduction of the Boyer-Finley 
equation \cite{FKS02}, etc.  

The work of Ferapontov, Korotkin and 
Shramchenko \cite{FKS02} is particularly suggestive, 
because their method is exactly based on 
L\"owner-type equations and some related equations.  
They indeed used those equations to apply Tsarev's 
generalized hodograph method \cite{Tsarev85}.  
On the other hand, since they do not use 
Lax equations, one cannot readily see how to generalize 
their results to higher flows of an underlying hierarchy.  
Therefore it is essential to understand their method 
in the perspective of Lax equations.  

Bearing these issues in mind, let us briefly 
look into rational reductions of the universal 
Whitham hierarchy.  Let us mention that 
the following consideration is more or less parallel 
to the approach that Gibbons et al. followed 
in the case of the Benney equations 
\cite{GT96,GT99,YG00}.

\paragraph*{Setup of rational reduction}
In a rational reduction, we assume that there is 
a rational function 
\beqnn
  E(p) = p^{k_0} + \sum_{n=2}^{k_0}a_{0n}p^{k_0-n} 
         + \sum_{\alpha=1}^N\sum_{n=1}^{k_\alpha} 
           \frac{a_{\alpha n}}{(p-q_\alpha)^n} 
\eeqnn
with poles at $p = \infty,q_1,\ldots,q_N$ such that 
$z_0(p)$ and $z_\alpha(p)$, $\alpha = 1,\ldots,N$, 
are given by Laurent expansion of fractional powers 
of $E(p)$ as 
\beqnn
\begin{aligned}
z_0(p) &= E(p)^{1/k_0} 
  &\quad (\mbox{Laurent expansion at $p=\infty$})\\
z_\alpha(p) &= E(p)^{1/k_\alpha} 
  &\quad (\mbox{Laurent expansion at $p=q_\alpha$}). 
\end{aligned}
\eeqnn
Actually, $E(p)$ can have extra (dynamical) poles 
other than $\infty$ and $q_\alpha$'s.  
This is indeed the case for, e.g., 
the Zakharov reductions of the Benney equations 
and the hydrodynamic reduction of the Boyer-Finley 
equations.  In those reductions, $E(p)$ takes 
such a form as 
\beqnn
  E(p) = p + \sum_{k=1}^{M-1}\frac{a_k}{p-b_k}, 
\eeqnn
whereas the Benney equations and the Boyer-Finley 
equations are embedded into the one- and 
two-point universal Whitham hierarchies 
(in other words, the dispersionless KP and 
Toda hierarchy).  

In this setup, the critical points 
\beqnn
  p = p_j,\quad E'(p_j) = 0, 
  \quad j = 1,\ldots,M, 
\eeqnn
and the critical values 
\beqnn
  \lambda_j = E(p_j), \quad j = 1,\ldots,M, 
\eeqnn
of $E(p)$ play the role of driving forces 
and Riemann invariants. More precisely, 
if $E(p)$ is sufficiently general, 
the critical values $\lambda_j$'s can be used 
as full parameters (or ``moduli'') of $E(p)$.  
Thus $E(p)$ is understood to be a function 
$E(p,\bslambda)$ of $p$ and $\bslambda 
= (\lambda_1,\ldots,\lambda_M)$.

\paragraph*{Hydrodynamic equations for $\lambda_j$}
The Lax equations (\ref{Lax-z(p)}) for the $z$-functions 
reduce to the Lax equations 
\beq
  \rd_{\alpha n}E(p) = \{\Omega_{\alpha n},E(p)\} 
  \quad (\alpha = 0,1,\ldots,N) 
\label{rat-Lax}
\eeq
for $E(p)$.  
Written more explicitly, these equations read 
\beqnn
  \rd_{\alpha n}E(p) 
  = \Omega_{\alpha n}'(p)\rd_{01}E(p) 
    - E'(p)\rd_{01}\Omega_{\alpha n}(p). 
\eeqnn
Letting $p = p_j$ in this equation 
yields the equation 
\beqnn
  \rd_{\alpha n}E(p)\Bigr|_{p=p_j} 
  = \Omega_{\alpha n}'(p_j)\rd_{01}E(p)\Bigr|_{p=p_j}. 
\eeqnn
On the other hand, by the chain rule, 
differentiating $\lambda_j = E(p_j)$ 
by $t_{\alpha n}$ gives the identity 
\beqnn
  \rd_{\alpha n}\lambda_j 
  = E'(p_j)\rd_{\alpha n}p_j 
    + \rd_{\alpha n}E(p)\Bigr|_{p=p_j} 
  = \rd_{\alpha n}E(p)\Bigr|_{p=p_j}. 
\eeqnn
Thus we obtain the diagonal hydrodynamic equations 
\beq
  \rd_{\alpha n}\lambda_j 
  = \chi_{\alpha n}(\bslambda)\rd_{01}\lambda_j 
  \quad (\alpha = 0,1,\ldots,N) 
\label{rat-dlambda}
\eeq
with characteristic speed 
\beqnn
  \chi_{\alpha n}(\bslambda) = \Omega_{\alpha n}'(p_j). 
\eeqnn
Note that $p_j$'s are now understood to be 
algebraic functions $p_j = p_j(\bslambda)$ 
defined by the equation $E'(p) = 0$. 
Thus (\ref{rat-dlambda}) may be thought of 
as a closed evolutionary system for $\bslambda$.  
As we shall show in a more general case, 
these reduced equations can be solved 
by the generalized hodograph method.

\paragraph*{L\"owner-type equations for $E(p)$}
We now consider $E(p)$ to be a function of 
$p$ and $\bslambda = \bslambda(\bst)$, 
and use the chain rule to rewrite 
the Lax equations (\ref{rat-Lax}).  
Both hand sides the Lax equations 
can be thereby expressed as 
\beqnn
  \mathrm{LHS} 
  = \sum_{j=1}^M 
    \frac{\rd E(p)}{\rd\lambda_j}\rd_{\alpha n}\lambda_j 
  = \sum_{j=1}^M 
    \frac{\rd E(p)}{\rd\lambda_j}
    \Omega_{\alpha n}'(p_j)\rd_{01}\lambda_j
\eeqnn
and 
\beqnn
  \mathrm{RHS}
  = \sum_{j=1}^M \left(
      \Omega_{\alpha n}'(p)\frac{\rd E(p)}{\rd\lambda_j} 
      - E'(p)\frac{\rd\Omega_{\alpha n}(p)}{\rd\lambda_j} 
    \right)\rd_{01}\lambda_j, 
\eeqnn
where we have used (\ref{rat-dlambda}) as well.  
Thus the Lax equations reduce to 
\beqnn
  \sum_{j=1}^M \left( 
    (\Omega_{\alpha n}'(p)-\Omega_{\alpha n}'(p_j))
    \frac{\rd E(p)}{\rd\lambda_j} 
    - E'(p)\frac{\rd\Omega_{\alpha n}(p)}{\rd\lambda_j} 
  \right)\rd_{01}\lambda_j 
  = 0. 
\eeqnn
Consequently, if $E(p) = E(p,\lambda)$ 
satisfies the equation
\beq
    (\Omega_{\alpha n}'(p)-\Omega_{\alpha n}'(p_j))
    \frac{\rd E(p)}{\rd\lambda_j} 
    - E'(p)\frac{\rd\Omega_{\alpha n}(p)}{\rd\lambda_j} 
  = 0,  
\label{rat-gen-Loewner}
\eeq
then for any solution $\bslambda = \bslambda(\bst)$  
of (\ref{rat-dlambda}), 
$E(p)\Bigr|_{\bslambda=\bslambda(\bst)}$ 
gives a solution of the Lax equation.  

Let us examine (\ref{rat-gen-Loewner}) 
in more detail.  When $\alpha = 0$ and $n = 1$, 
this equation is a trivial identity. 
The lowest nontrivial one is the case where 
$\alpha = 0$ and $n = 2$. In this case, 
since $\Omega_{02}(p) = p^2 + 2u_{02}$, 
(\ref{rat-gen-Loewner}) reduces to 
\beqnn
  (p - p_j)\frac{\rd E(p)}{\rd\lambda_j} 
  - E'(p)\frac{\rd u_{02}}{\rd\lambda_j} = 0 
\eeqnn
or, equivalently, 
\beq
  \frac{\rd E(p)}{\rd\lambda_j} 
  = \frac{E'(p)}{p - p_j}\frac{\rd u_{02}}{\rd\lambda_j}.
\label{rat-Loewner}
\eeq
This is exactly a L\"owner-type equation. 
We can readily derive the equations 
\beq
  \frac{\rd z_\alpha(p)}{\rd\lambda_j} 
  = \frac{z_\alpha'(p)}{p - p_j}
    \frac{\rd u_{02}}{\rd\lambda_j} 
\eeq
for the $z_\alpha(p)$'s, because the $z$-functions 
are given by fractional powers of $E(p)$.  
The $p$-functions, in turn, satisfy 
the dual equations 
\beq
  \frac{\rd p_\alpha(z)}{\rd\lambda_j} 
  = \frac{1}{p_j - p_\alpha(z)}
    \frac{\rd u_{02}}{\rd\lambda_j}. 
\eeq

As regards the case for $n > 2$, 
(\ref{rat-gen-Loewner}) turn out to be redundant, 
namely, automatically satisfied 
if (\ref{rat-Loewner}) is satisfied.  
This fact can be explained in a more general form; 
we shall return to this issue in Section 6.2.

\section{Multi-variable reductions}

In view of the foregoing interpretation of 
rational reductions, it is now rather 
straightforward to find a correct formulation 
of general multi-variable reductions.  
In this section, we present this formulation 
and its implications.  After all, this reduction 
procedure is nothing but the ``diagonal reduction'' 
in the sense of Guil et al. \cite{GMMA03}. 
We, however, attempt to reformulate it 
along the lines that we have pursued 
in the case of one-variable reduction.

\subsection{L\"owner-type equations for $p_\alpha(z)$ and $z_\alpha(p)$}

In an $M$-variable reduction, the fundamental 
dynamical variables $u_{\alpha n}$ ($n = 0,1,\ldots$, 
$u_{\alpha 0} = r_\alpha$) and $q_\alpha$ 
are assumed to be functions of $M$-dimensional 
reduced dynamical variables $\bslambda 
= (\lambda_1,\ldots,\lambda_M)$.  
The reduced dynamical variables, in turn, 
depend on $\bst$ as $\bslambda = \bslambda(\bst)$ 
and eventually satisfy a set of diagonal 
hydrodynamic evolution equations.  

Accordingly, the $p$-functions $p_\alpha(z)$ 
and the $z$-functions $z_\alpha(p)$, 
$\alpha = 0,1,\ldots,N$, are functions 
of $p$ and $\bslambda$, 
\beqnn
  z_\alpha(p) = z_\alpha(p,\bslambda), \quad 
  p_\alpha(z) = p_\alpha(z,\bslambda). 
\eeqnn
We assume that these functions satisfy 
the L\"owner-type equations 
\beq
  \frac{\rd p_\alpha(z)}{\rd\lambda_j} 
  = \frac{1}{U_j - p_\alpha(z)}\frac{\rd u_{02}}{\rd\lambda_j}, 
  \quad j = 1,\ldots,M, 
\label{M-Loewner-p(z)}
\eeq
or the dual equations
\beq
  \frac{\rd z_\alpha(p)}{\rd\lambda_j}
  = \frac{z_\alpha'(p)}{p - U_j}\frac{\rd u_{02}}{\rd\lambda_j}, 
  \quad j=1,\ldots,M, 
\label{M-Loewner-z(p)}
\eeq
for a given set of driving forces $U_j = U_j(\bslambda)$ 
and the auxiliary function $u_{02} = u_{02}(\bslambda)$. 
Unlike the one-variable case, these auxiliary functions 
have to satisfy a set of integrability conditions.  
We shall present these conditions later on 
when we consider the hodograph method.  

As in the case of one-variable reduction, 
the dual equations (\ref{M-Loewner-z(p)}) 
imply that the $z$-functions are functionally 
related with each other by functions 
$f_\alpha(z)$ of one variable $z$ as 
(\ref{zalpha(p)-z0(p)}) shows.  
Though Guil et al. \cite{GMMA03} further assumed 
the special form (\ref{GMMA03-assume}), 
the following consideration is not limited 
to that case.  

As a consequence of these L\"owner-type equations, 
$z_\alpha(p)$'s turn out to have critical points 
at $p = U_j$, namely, 
\beq
  z'_\alpha(U_j) = 0. 
\eeq
This is an immediate consequence of the structure 
of (\ref{M-Loewner-z(p)}).  Since the left hand side 
has no singularity at $p = U_j$, the pole of 
$1/(p-U_j)$ at $p = U_j$ has to be canceled by 
a zero of $z_\alpha'(p)$.  This, however, does not 
imply that the critical values $z_\alpha(U_j)$ 
coincide with $\lambda_j$'s.  This is in accord 
with the general fact that the choice of 
Riemann invariants is not unique 
but allows large arbitrariness.  A standard way 
will be to choose the critical values $z_0(U_j)$ 
of $z_0(z)$ as $\lambda_j$'s;  the critical values 
$z_\alpha(U_j)$ of the other $z$-functions 
are then functionally related to $z_0(U_j)$ 
as (\ref{zalpha(p)-z0(p)}) implies.

\subsection{Hydrodynamic equations for $\lambda_j$}

In view of the results on one-variable and 
rational reductions, it seems plausible that 
the reduced variables $\bslambda = \bslambda(\bst)$ 
satisfy diagonal hydrodynamic equations of the form 
\beq
  \rd_{\alpha n}\lambda_j 
  = \chi_{\alpha nj}(\bslambda)\rd_{01}\lambda_j 
\label{M-dlambda}
\eeq
with characteristic speeds
\beqnn
  \chi_{\alpha nj}(\bslambda) 
  = \Omega_{\alpha n}'(U_j). 
\eeqnn
The relation between the Lax equations 
(\ref{Lax-z(p)}) and these equations, 
however, is more delicate than 
in the case of rational reductions, 
because $\lambda_j$'s in the present setup 
are {\it not} assumed to be given by 
the critical values of the $z$-functions.  
Therefore we cannot derive (\ref{M-dlambda}) 
by simply letting $p = U_j$ in the Lax equations. 
Nevertheless, we can confirm that 
(\ref{M-dlambda}) are correct equations 
to be satisfied by $\bslambda = \bslambda(\bst)$. 

\begin{theorem}\label{th:dlambda-Lax}
If (\ref{M-Loewner-p(z)}) (or (\ref{M-Loewner-z(p)})) 
and (\ref{M-dlambda}) are satisfied, 
then the Lax equations (\ref{Lax-z(p)}) 
are also satisfied. 
\end{theorem}

Although this theorem can also be deduced from 
some other results that we shall show later on 
(Theorem \ref{th:M-integrability}
and Theorem \ref{th:S-function}) , 
we dare to present a direct proof here.  
The outline of the proof is parallel to 
the case of rational reductions.  
A clue is the following 

\begin{lemma}\label{lem:dOmega/dlambda}
If $p_\alpha(z)$'s satisfy (\ref{M-Loewner-p(z)}), 
then $\Omega_{\alpha n}(p)$'s satisfy the identities 
\beq
  \frac{\rd\Omega_{\alpha n}(p)}{\rd\lambda_j} 
  = \frac{\Omega_{\alpha n}'(p)-\Omega_{\alpha n}'(U_j)}{p-U_j} 
    \frac{\rd u_{02}}{\rd\lambda_j}. 
\label{dOmega/dlambda}
\eeq
\end{lemma}

\paragraph*{Proof}
Let us first consider the case where $\alpha = 0$. 
We differentiate the generating function (\ref{Faber}) 
of $\Omega_{0n}(p)$'s by $\lambda_j$ and 
use (\ref{M-Loewner-p(z)}).  This yields 
the identity 
\beqnn
\begin{aligned}
- \sum_{n=1}^\infty \frac{z^{-n}}{n} 
      \frac{\rd\Omega_{0n}(p)}{\rd\lambda_j} 
&= \frac{1}{p_0(z)-p}\frac{\rd p_0(z)}{\rd\lambda_j} \\
&= - \frac{1}{(p-p_0(z))(U_j-p_0(z))}
     \frac{\rd u_{02}}{\rd\lambda_j}. 
\end{aligned}
\eeqnn
On the other hand, the kernel formula (\ref{kernel}) 
implies another identity of the form 
\beqnn
\begin{aligned}
- \sum_{n=1}^\infty \frac{z^{-n}}{n} 
    (\Omega_{0n}'(p) - \Omega_{0n}'(U_j)) 
&= \frac{1}{p-p_0(z)} - \frac{1}{U_j-p_0(z)} \\
&= \frac{U_j-p}{(p-p_0(z))(U_j-p_0(z))}. 
\end{aligned}
\eeqnn
By comparing these identities, we find 
the identity 
\beqnn
  \sum_{n=1}^\infty \frac{z^{-n}}{n} 
    \frac{\rd\Omega_{0n}(p)}{\rd\lambda_j} 
= \sum_{n=1}^\infty \frac{z^{-n}}{n} 
    \frac{\Omega_{0n}'(p)-\Omega_{\alpha 0}'(U_j)}{p-U_j}, 
\eeqnn
which implies that (\ref{dOmega/dlambda}) 
holds for $\alpha = 0$.  In much the same way
using the second kernel formula (\ref{kernel2}),
we can derive (\ref{dOmega/dlambda}) 
for $\alpha = 1,\ldots,N$.  \qed

\paragraph*{Proof of Theorem \ref{th:dlambda-Lax}}
The reasoning in the case of rational reductions 
also works in this case with slightest modification. 
Firstly, using (\ref{M-dlambda}) and the chain rule, 
we can rewrite the Lax equations (\ref{Lax-z(p)}) as 
\beqnn
  \sum_{j=1}^M \left( 
    (\Omega_{\alpha n}'(p)-\Omega_{\alpha n}'(U_j))
    \frac{\rd z_\beta(p)}{\rd\lambda_j} 
    - z_\beta'(p)\frac{\rd\Omega_{\alpha n}(p)}{\rd\lambda_j} 
  \right)\rd_{01}\lambda_j 
  = 0. 
\eeqnn
On the other hand, combining the equations 
(\ref{M-Loewner-z(p)}) for $z_\beta(p)$ 
with the identity (\ref{dOmega/dlambda}) of 
the lemma yields the equations 
\beq
  \frac{\rd z_\beta(p)}{\rd\lambda_j} 
  = \frac{z_\beta'(p)}{\Omega_{\alpha n}'(p)-\Omega_{\alpha n}'(U_j)} 
    \frac{\rd\Omega_{\alpha n}(p)}{\rd\lambda_j}, 
\label{M-gen-Loewner}
\eeq
which implies that the last equations 
are indeed satisfied.  \qed

Let us note that the last equations 
(\ref{M-gen-Loewner}) amount to 
(\ref{rat-gen-Loewner}) in the previous section.  
As mentioned therein, (\ref{rat-gen-Loewner}) 
contains the L\"owner-type equations 
(\ref{rat-Loewner}) as a special case 
with $\alpha = 0$ and $n = 2$.  
In the present setup, (\ref{M-Loewner-z(p)}) 
is a special case of (\ref{M-gen-Loewner}) 
with $\alpha = 0$ and $n = 2$. 
By the way, Lemma\ref{lem:dOmega/dlambda} 
(or its proof) says that (\ref{M-gen-Loewner}) 
is a consequence of (\ref{M-Loewner-z(p)}).  
This explains why (\ref{rat-gen-Loewner}) 
are ``redundant'' in the setup of the last section.

\subsection{Hodograph method}

The hydrodynamic equations (\ref{M-dlambda}) 
can be solved by Tsarev's hodograph method \cite{Tsarev85}.  
As it turns out below, this is a straightforward 
generalization of the framework developed 
by Gibbons and Tsarev \cite{GT96,GT99} 
for reductions of the Benney equations.  

In the multi-variable case, 
(\ref{M-Loewner-p(z)}) and (\ref{M-Loewner-z(p)}) 
have to satisfy a set of integrability conditions.  
As regards (\ref{M-Loewner-p(z)}), 
the integrability conditions can be derived 
by eliminating the $\bslambda$-derivatives 
of $p_\alpha(z)$ from the identity 
\beqnn
  \frac{\rd}{\rd\lambda_j}
  \left(\frac{1}{U_k-p_\alpha(z)}\frac{\rd u_{02}}{\rd\lambda_k}\right) 
= \frac{\rd}{\rd\lambda_k}
  \left(\frac{1}{U_j-p_\alpha(z)}\frac{\rd u_{02}}{\rd\lambda_j}\right). 
\eeqnn
After some algebra, these conditions 
reduce to the equations 
\beq
\begin{aligned}
\frac{\rd U_k}{\rd\lambda_j} 
  &= \frac{1}{U_j-U_k}\frac{\rd u_{02}}{\rd\lambda_j}, \\
\frac{\rd^2u_{02}}{\rd\lambda_j\rd\lambda_k} 
  &= \frac{2}{(U_j-U_k)^2}
     \frac{\rd u_{02}}{\rd\lambda_j}
     \frac{\rd u_{02}}{\rd\lambda_k}, 
\end{aligned}
\label{Gibbons-Tsarev}
\eeq
which take exactly the same form as 
the equations derived by Gibbons and Tsarev 
in the case of the Benney equations.  
The same equations can be derived from 
the integrability conditions 
\beq
\left[
  \frac{\rd}{\rd\lambda_j} 
  - \frac{1}{p-U_j}\frac{\rd u_{02}}{\rd\lambda_j}\frac{\rd}{\rd p},\, 
  \frac{\rd}{\rd\lambda_k}
  - \frac{1}{p-U_k}\frac{\rd u_{02}}{\rd\lambda_k}\frac{\rd}{\rd p}
\right] 
= 0 
\label{Loewner-integrable}
\eeq
of the dual equations (\ref{M-Loewner-z(p)}) as well. 

An important consequence of these equations is the following

\begin{lemma}
The characteristic speeds $\chi_{\alpha nj} 
= \chi_{\alpha nj}(\bslambda)$ of (\ref{M-dlambda}) 
satisfy the equations 
\beq
  \frac{\rd\chi_{\alpha nk}}{\rd\lambda_j} 
  = (\chi_{\alpha nj} - \chi_{\alpha nk})V_{jk}, 
\label{dchi/dlambda}
\eeq
where 
\beq
  V_{jk} = \frac{1}{(U_j-U_k)^2}\frac{\rd u_{02}}{\rd\lambda_j}. 
\eeq
\end{lemma}

\paragraph*{Proof}
Let us first consider the case where $\alpha = 0$. 
Substituting $q = U_k$ in the kernel formula (\ref{kernel}), 
yields the identity 
\beqnn
  - \sum_{n=1}^\infty \frac{z^{-n}}{n}\chi_{0nk}
  = \frac{1}{U_k-p_0(z)}. 
\eeqnn
We now differentiate both hand sides by $\lambda_j$. 
Using (\ref{M-Loewner-p(z)}) and (\ref{Gibbons-Tsarev}), 
we can rewrite the outcome as 
\beqnn
\begin{aligned}
- \sum_{n=1}^\infty 
   \frac{z^{-n}}{n}\frac{\rd\chi_{0nk}}{\rd\lambda_j} 
&= - \frac{1}{(U_k-p_0(z))^2}
     \left(\frac{\rd U_k}{\rd\lambda_j} 
       - \frac{\rd p_0(z)}{\rd\lambda_j}\right) \\
&= - \frac{1}{(U_k-p_0(z))^2}
     \left(\frac{1}{U_j-U_k} - \frac{1}{U_j-p_0(z)}\right)
     \frac{\rd u_{02}}{\rd\lambda_j} \\
&= - \frac{1}{(U_k-p_0(z))(U_j-p_0(z))(U_j-U_k)}
     \frac{\rd u_{02}}{\rd\lambda_j} \\
&=  \left(\frac{1}{U_j-p_0(z)}-\frac{1}{U_k-p_0(z)}\right)
    \frac{1}{(U_j-U_k)^2}\frac{\rd u_{02}}{\rd\lambda_j} \\
&= - \sum_{n=1}^\infty 
     \frac{z^{-n}}{n}(\chi_{\alpha nj}-\chi_{\alpha nk})
     \frac{1}{(U_j-U_k)^2}\frac{\rd u_{02}}{\rd\lambda_j}. 
\end{aligned}
\eeqnn
This shows that (\ref{dchi/dlambda}) 
are indeed satisfied for $\alpha = 0$.  
We can confirm (\ref{dchi/dlambda}) 
for $\alpha = 1,\ldots,N$ in the same way, 
now using the second kernel formula (\ref{kernel2}). 
\qed

We can now formulate the hodograph method for 
(\ref{M-dlambda}) as follows. 

\begin{theorem}
Let $F_j = F_j(\bslambda)$, $j = 1,\ldots,M$, 
be a set of functions of $\bslambda$ that satisfy 
the equations 
\beq
  \frac{\rd F_k}{\rd\lambda_j} = (F_j - F_k)V_{jk}, 
\label{rdF/dlambda}
\eeq
and $\bslambda = \bslambda(\bst)$ a solution of 
the hodograph equations 
\beq
  \sum_{n=1}^\infty t_{0n}\chi_{0nj}(\bslambda) 
  + \sum_{\alpha=1}^N\sum_{n=0}^\infty 
      t_{\alpha n}\chi_{\alpha nj}(\bslambda) 
  = F_j(\bslambda), 
  \quad j = 1,\ldots,N. 
\label{M-hodograph}
\eeq
Further assume that the regularity conditions 
\beq
  \sum_{n=1}^\infty 
    t_{0n}\frac{\rd\chi_{0nj}(\bslambda)}{\rd\lambda_j}
  + \sum_{\alpha=1}^N\sum_{n=0}^\infty 
      t_{\alpha n}
      \frac{\rd\chi_{\alpha nj}(\bslambda)}{\rd\lambda_j} 
  \not= \frac{\rd F_j(\bslambda)}{\rd\lambda_j}, 
  \quad j = 1,\ldots,N, 
\eeq
hold for $\bslambda = \bslambda(\bst)$.  
Then $\bslambda = \bslambda(\bst)$ satisfies 
the hydrodynamic equations (\ref{M-dlambda}). 
\end{theorem}

\paragraph*{Proof}
We differentiate both hand sides of 
(\ref{M-hodograph}) by $t_{\alpha n}$.  
By the chain rule, this yields the equations 
\beqnn
\chi_{\alpha nj}
+ \sum_{k=1}^M\left(
    \sum_{m=1}^\infty 
      t_{0m}\frac{\rd\chi_{0mj}}{\rd\lambda_k}
    + \sum_{\beta=1}^N\sum_{m=0} 
      t_{\beta m}\frac{\rd\chi_{\beta mj}}{\rd\lambda_k}
  \right)\rd_{\alpha n}\lambda_k 
= \sum_{k=1}^M
  \frac{\rd F_j}{\rd\lambda_k}\rd_{\alpha n}\lambda_k, 
\eeqnn
hence 
\beqnn
\chi_{\alpha nj} 
= \sum_{k=1}^M\left( 
     \frac{\rd F_j}{\rd\lambda_k}
     - \sum_{m=1}^\infty     
        t_{0m}\frac{\rd\chi_{0mj}}{\rd\lambda_k}
     - \sum_{\beta=1}^N\sum_{m=0}^\infty 
        t_{\beta m}\frac{\rd\chi_{\beta mj}}{\rd\lambda_k}
  \right)\rd_{\alpha n}\lambda_k. 
\eeqnn
Let us examine the quantity inside the parenthesis 
on the right hand side.  If $k \not= j$, we can use 
(\ref{dchi/dlambda}) and (\ref{rdF/dlambda}) 
to rewrite this quantity as 
\beqnn
\begin{aligned}
& \frac{\rd F_j}{\rd\lambda_k}
  - \sum_{m=1}^\infty     
      t_{0m}\frac{\rd\chi_{0mj}}{\rd\lambda_k}
  - \sum_{\beta=1}^N\sum_{m=0}^\infty 
      t_{\beta m}\frac{\rd\chi_{\beta mj}}{\rd\lambda_k}\\
&= (F_k-F_j)V_{kj} 
   - \sum_{m=1}^\infty 
      t_{0m}(\chi_{0mk}-\chi_{0mj})V_{kj}
   - \sum_{\beta=1}^N\sum_{m=0}^\infty 
      t_{\beta m}(\chi_{\beta mk}-\chi_{\beta mj})V_{kj} \\
&= \left(F_k - \sum_{m=1}^\infty t_{0m}\chi_{0mk}
     - \sum_{\beta=1}^N\sum_{m=0}^\infty t_{\beta m}\chi_{\beta mk}
     - (\mbox{$k$ replaced with $j$}) 
   \right)V_{kj} 
\end{aligned}
\eeqnn
which vanishes by (\ref{M-hodograph}). 
Thus the last equation simplifies as 
\beqnn
\chi_{\alpha nj} 
= \left(
     \frac{\rd F_j}{\rd\lambda_j}
     - \sum_{m=1}^\infty     
        t_{0m}\frac{\rd\chi_{0mj}}{\rd\lambda_j}
     - \sum_{\beta=1}^N\sum_{m=0}^\infty 
        t_{\beta m}\frac{\rd\chi_{\beta mj}}{\rd\lambda_j}
  \right)\rd_{\alpha n}\lambda_j. 
\eeqnn
Moreover, if $\alpha = 0$ and $n = 2$, 
this equation reduces to 
\beqnn
1 = \left(
     \frac{\rd F_j}{\rd\lambda_j}
     - \sum_{m=1}^\infty     
        t_{0m}\frac{\rd\chi_{0mj}}{\rd\lambda_j}
     - \sum_{\beta=1}^N\sum_{m=0}^\infty 
        t_{\beta m}\frac{\rd\chi_{\beta mj}}{\rd\lambda_j}
    \right)\rd_{\alpha 0}\lambda_j. 
\eeqnn
If we multiply the last equation by $\chi_{\alpha nj}$ 
and subtract it from the previous one, 
the outcome is the equation 
\beqnn
\begin{aligned}
0 &= \left(
     \frac{\rd F_j}{\rd\lambda_j}
     - \sum_{m=1}^\infty     
        t_{0m}\frac{\rd\chi_{0mj}}{\rd\lambda_j}
     - \sum_{\beta=1}^N\sum_{m=0}^\infty 
        t_{\beta m}\frac{\rd\chi_{\beta mj}}{\rd\lambda_j}
    \right) \times \\
  &\quad \times 
    (\rd_{\alpha n}\lambda_j 
     - \chi_{\alpha nj}\rd_{\alpha 01}\lambda_j). 
\end{aligned}
\eeqnn
By the regularity condition, we can drop 
the prefactor of $\chi_{\alpha n}\lambda_j 
- \chi_{\alpha nj}\rd_{\alpha 01}\lambda_j$ 
and obtain the hydrodynamic equations (\ref{M-dlambda}).  
\qed

We are thus eventually left with the problem of 
finding $F_j$'s that satisfy (\ref{rdF/dlambda}).  
Such functions are given by contour integrals 
of the form 
\beq
  F_j = \sum_{\alpha=0}^N \oint_{C_\alpha}\frac{dp}{2\pi i} 
        \frac{G_\alpha(z_\alpha(p))}{(p - U_j)^2}, 
\eeq
where $G_\alpha(z)$ is an arbitrary holomorphic function 
of one variable $z$ defined on the range of 
the map $p \mapsto z_\alpha(p)$, 
and $C_\alpha$ is a closed curve (or cycle) 
in the domain of $z_\alpha(p)$.  
It is not difficult to show that these $F_j$'s 
satisfy (\ref{rdF/dlambda}) as a consequence 
of (\ref{M-Loewner-z(p)}).   Note that 
the characteristic speeds $\chi_{\alpha nj}$ 
themselves have such a contour integral 
representation as 
\beq
  \chi_{\alpha nj} 
  = \oint\frac{dp}{2\pi i}
    \frac{z_\alpha(p)^n}{(p - U_j)^2}, 
\eeq
where the path of integral is a small circle 
encircling the point $p = \infty$ 
($\alpha = 0$) or $p = q_\alpha$ 
($\alpha = 1,\ldots,N$).

\subsection{Existence of $F$-function}

Theorem \ref{th:integrability} and its proof 
can be generalized to the present setup 
without substantial modifications.  

\begin{theorem}\label{th:M-integrability}
The integrability conditions (\ref{ddF-integrable}) 
of (\ref{ddF}) are satisfied in the foregoing setup 
of multi-variable reduction.  The $F$-function 
$\mathcal{F} = \mathcal{F}(\bst)$ thus defined 
by (\ref{ddF}) gives a solution of 
the dispersionless Hirota equations (\ref{dHirota}). 
\end{theorem}

\paragraph*{Proof}
We can proceed just as in the proof of 
Theorem \ref{th:integrability}.  By applying 
$\hat{D}_\gamma(z)$ to the generating functions 
(\ref{Grunsky}), we obtain the generating functions 
\beqnn
  \frac{\hat{D}_\gamma(z_3)(p_\alpha(z_1)-p_\beta(z_2))}
       {p_\alpha(z_1)-p_\beta(z_2)} 
= - \sum_{l,m,n}z_1^{-l}z_2^{-m}z_3^{-n} 
    \hat{\rd}_{\gamma n}b_{\alpha l\beta m} 
\eeqnn
of derivatives of the Grunsky coefficients. 
By the kernel formulas (\ref{kernel}) and 
(\ref{kernel2}), the hydrodynamic equations 
(\ref{M-dlambda}) can be cast into 
the generating functional form 
\beqnn
  \hat{D}_\alpha(z)\lambda_j 
  = \frac{\rd_{01}\lambda_j}{p_\alpha(z)-U_j}. 
\eeqnn
We can thereby rewrite the foregoing 
generating function as 
\beqnn
\begin{aligned}
& \frac{\hat{D}_\gamma(z_3)(p_\alpha(z_1)-p_\beta(z_2))}
       {p_\alpha(z_1)-p_\beta(z_2)} \\
&= - \sum_{j=1}^M \frac{\rd_{01}\lambda_j}
     {(U_j-p_\alpha(z_1))(U_j-p_\beta(z_2))(U_j-p_\gamma(z_3))}
     \frac{\rd u_{02}}{\rd\lambda_j}. 
\end{aligned}
\eeqnn
This implies the functional identity 
\beqnn
  \frac{\hat{D}_\gamma(z_3)(p_\alpha(z_1)-p_\beta(z_2))}
       {p_\alpha(z_1)-p_\beta(z_2)} 
= \frac{\hat{D}_\alpha(z_1)(p_\gamma(z_3)-p_\beta(z_2))}
       {p_\gamma(z_3)-p_\beta(z_2)}. 
\eeqnn
The integrability conditions (\ref{ddF-integrable}) 
follow from this identity immediately.  
The rest of the statement of the theorem 
is a consequence of the comments 
in the end of Section 2.4.  
\qed

As a byproduct of this proof, we obtain 
the following generalization of (\ref{dddF}) 
to multi-variable reductions: 
\beq
\begin{aligned}
& \sum_{l,m,n}z_1^{-l}z_2^{-m}z_3^{-n}
  \hat{\rd}_{\alpha l}\hat{\rd}_{\beta m}\hat{\rd}_{\gamma n}\mathcal{F}\\
&= - \sum_{j=1}^M 
     \frac{\rd_{01}\lambda_j}
     {(U_j-p_\alpha(z_1))(U_j-p_\beta(z_2))(U_j-p_\gamma(z_3))} 
     \frac{\rd u_{02}}{\rd\lambda_j}. 
\end{aligned}
\label{M-dddF}
\eeq

\section{$S$-functions in multi-variable reductions}

As an application of the foregoing formulation 
of multi-variable reductions, we now reconsider 
the construction of $S$-functions by 
Guil et al. \cite{GMMA03}.  As it turns out below, 
some part of their construction can be made 
more transparent with the aid of the identities 
(\ref{dOmega/dlambda}) of Lemma\ref{lem:dOmega/dlambda}.

The construction of $S$-functions by Guil. et al. 
is based on hodograph solutions of 
the hydrodynamic equations (\ref{M-dlambda}).  
Given such a solution along with the $p$-functions 
satisfying (\ref{M-Loewner-p(z)}), 
they construct the $S$-function in such a form as 
\beq
  \mathcal{S}_\beta(p) 
  = \sum_{n=1}^\infty t_{0n}\Omega_{0n}(p) 
    + \sum_{\alpha=1}^N\sum_{n=0}^\infty 
        t_{\alpha n}\Omega_{\alpha n}(p) 
    + \mathcal{S}_{\beta -}(p) 
\eeq
or, equivalently, 
\beq
  S_\beta(z)
  = \sum_{n=1}^\infty t_{0n}\Omega_{0n}(p_\beta(z)) 
    + \sum_{\alpha=1}^N\sum_{n=0}^\infty 
        t_{\alpha n}\Omega_{\alpha n}(p_\beta(z)) 
    + \mathcal{S}_{\beta -}(p_\beta(z)) 
\eeq
where $\mathcal{S}_{\beta -}(p)$ are 
required to satisfy the equations 
\beq
  \left(\frac{\rd}{\rd\lambda_j} 
    - \frac{1}{p-U_j}\frac{\rd u_{02}}{\rd\lambda_j}
      \frac{\rd}{\rd p}
  \right)\mathcal{S}_{\beta -}(p) 
  = - \frac{F_j}{p-U_j}\frac{\rd u_{02}}{\rd\lambda_j}. 
\label{Sminus}
\eeq
Since the differential operators on the left hand side 
are commutative, see (\ref{Loewner-integrable}), 
the integrability conditions of these 
{\it inhomogeneous} L\"owner equations 
are given by 
\beqnn
\begin{aligned}
& \left(\frac{\rd}{\rd\lambda_j}
    - \frac{1}{p-U_j}\frac{\rd u_{02}}{\rd\lambda_j}
      \frac{\rd}{\rd p}\right)
  \left(\frac{F_k}{p-U_k}\frac{\rd u_{02}}{\rd\lambda_k}\right) \\
&= \left(\frac{\rd}{\rd\lambda_k}
     - \frac{1}{p-U_k}\frac{\rd u_{02}}{\rd\lambda_k}
       \frac{\rd}{\rd p}\right)
   \left(\frac{F_j}{p-U_j}\frac{\rd u_{02}}{\rd\lambda_j}\right). 
\end{aligned}
\eeqnn
By straightforward calculations, one can see 
that these conditions are equivalent to 
the equations (\ref{rdF/dlambda}) for $F_j$'s.  
Thus the existence of a solution to (\ref{Sminus}) 
is ensured in the setup of the hodograph solution. 
With these definitions, the main part of the results 
of Guil et al. can be stated as follows: 

\begin{theorem}[Guil, Ma\~{n}as, Mart\'{\i}nez Alonso \cite{GMMA03}]
\label{th:S-function}
The $S$-functions $S_\beta(z)$, $\beta = 0,1,\ldots,N$, 
satisfy the Hamilton-Jacobi equations (\ref{Hamilton-Jacobi}). 
\end{theorem}

Let us prove this theorem using our tools. 
Firstly, by the chain rule, 
the $t_{\alpha n}$-derivative of $S_\beta(z)$ 
can be expanded as 
\beq
\begin{aligned}
\rd_{\alpha n}S_\beta(z) 
&= \Omega_{\alpha n}(p_\beta(z)) \\
&\quad 
  + \sum_{j=1}^M \sum_{\gamma,m}
      t_{\gamma m}\left( 
        \Omega_{\gamma m}'(p_\beta(z))
        \frac{\rd p_\beta(z)}{\rd\lambda_j} 
        + \frac{\rd\Omega_{\gamma m}(p)}{\rd\lambda_j}
          \Bigr|_{p=p_\beta(z)} 
      \right)\rd_{\alpha n}\lambda_j \\
&\quad 
  + \sum_{j=1}^M \left(
      \mathcal{S}_{\beta -}'(p_\beta(z))
      \frac{\rd p_\beta(z)}{\rd\lambda_j} 
      + \frac{\rd\mathcal{S}_{\beta -}(p)}{\rd\lambda_j}
        \Bigr|_{p=p_\beta(z)} 
    \right)\rd_{\alpha n}\lambda_j, 
\end{aligned}
\label{GMMA03-HJ-pr1}
\eeq
where we have used the abbreviated notation 
\beqnn
  \sum_{\gamma m}A_{\gamma m}
  = \sum_{m=1}^\infty A_{0m} 
    + \sum_{\gamma=1}^N\sum_{m=0}^\infty A_{\gamma m}.
\eeqnn
Using the L\"owner-like equations (\ref{M-Loewner-p(z)}) 
and the identity (\ref{dOmega/dlambda}) of 
Lemma\ref{lem:dOmega/dlambda}, we can calculate 
the quantity in the first parenthesis 
on the right hand side of (\ref{GMMA03-HJ-pr1})
\beqnn
\begin{aligned}
&   \Omega_{\gamma m}'(p_\beta(z))
    \frac{\rd p_\beta(z)}{\rd\lambda_j} 
  + \frac{\rd\Omega_{\gamma m}(p)}{\rd\lambda_j}
    \Bigr|_{p=p_\beta(z)} \\
&=  \frac{\Omega_{\gamma m}'(p_\beta(z))}{U-p_\beta(z)}
    \frac{\rd u_{02}}{\rd\lambda_j} 
  + \frac{\Omega_{\gamma m}'(p_\beta(z))-\Omega_{\gamma m}'(U_j)}
         {p_\beta(z)-U_j}
    \frac{\rd u_{02}}{\rd\lambda_j} \\
&= - \frac{\Omega_{\gamma m}'(U_j)}{p_\beta(z)-U_j}
     \frac{\rd u_{02}}{\rd\lambda_j}.
\end{aligned}
\eeqnn
As regards the quantity in the second parenthesis, 
we use the equations (\ref{Sminus}) satisfied 
by $\mathcal{S}_{\beta -}(z)$ as 
\beqnn
\begin{aligned}
&   \mathcal{S}_{\beta -}'(p_\beta(z))
    \frac{\rd p_\beta(z)}{\rd\lambda_j} 
  + \frac{\rd\mathcal{S}_{\beta -}(p)}{\rd\lambda_j}
    \Bigr|_{p=p_\beta(z)} \\
&=  \frac{\mathcal{S}_{\beta -}'(p_\beta(z))}
         {U_j-p_\beta(z)}
    \frac{\rd u_{02}}{\rd\lambda_j} 
  + \frac{\rd\mathcal{S}_{\beta -}(p)}{\rd\lambda_j}
    \Bigr|_{p=p_\beta(z)} \\
&= \left(
     \frac{1}{U_j-p}\frac{\rd u_{02}}{\rd\lambda_j} 
     \frac{\rd\mathcal{S}_{\beta -}(p)}{\rd p}
     + \frac{\rd\mathcal{S}_{\beta -}(p)}{\rd\lambda_j}
   \right)\Bigr|_{p=p_\beta(z)}\\
&= - \frac{F_j}{p_\beta(z)-U_j}
     \frac{\rd u_{02}}{\rd\lambda_j}. 
\end{aligned}
\eeqnn
Consequently, (\ref{GMMA03-HJ-pr1}) 
turns into an equation of the form 
\beqnn
\rd_{\alpha n}S_\beta(z) 
= \Omega_{\alpha n}(p_\beta(z)) 
   + \sum_{j=1}^M 
       \left( \sum_{\gamma m}t_{\gamma m}
       \Omega_{\gamma m}'(U_j) 
       + F_j \right)
       \frac{\rd_{\alpha n}\lambda_j}{p_\beta(z)-U_j}
       \frac{\rd u_{02}}{\rd\lambda_j}. 
\eeqnn
Since each term of the sum on the right hand side 
vanishes by the hodograph equations 
(\ref{M-hodograph}), we have the equation 
\beqnn
  \rd_{\alpha n}S_\beta(z) = \Omega_{\alpha n}(p_\beta(z)). 
\eeqnn
In the case where $\alpha = 0$ and $n = 1$, 
this equation reduces to 
\beqnn
  \rd_{01}S_\beta(z) = p_\beta(z), 
\eeqnn
by which we can eliminate $p_\beta(z)$ from 
the last equation and obtain the Hamilton-Jacobi 
equations (\ref{Hamilton-Jacobi}). 

This completes the proof of the theorem.  
Note that using the identities 
(\ref{dOmega/dlambda}) makes the proof 
shorter and more understandable than 
the original proof of Guil et al.

\section{Conclusion}

We have thus seen that L\"owner-type equations 
play a fundamental role in finite variable reductions 
of the universal Whitham hierarchy of genus zero.  
The status of dispersionless Hirota equations therein 
is more subtle.  As regards the one-variable reduction, 
the dispersionless Hirota equations (\ref{dHirota}) 
are certainly a clue.  The generating functional form 
(\ref{log-dHirota}) of these equations enabled us 
to derive the L\"owner-type equations 
(\ref{Loewner-p(z)}) and (\ref{Loewner-z(p)}) 
directly from the assumption that all dynamical variables 
are functions of a single reduced variable $\lambda$. 
Unfortunately, this  method does not work for 
multi-variable reductions.  Therefore we were forced 
to start from (rather than derive) 
the L\"owner-type equations (\ref{M-Loewner-p(z)}) 
and (\ref{M-Loewner-z(p)}) and to confirm that 
they are a correct set of reduction conditions 
(Theorem \ref{th:dlambda-Lax}).   For both 
one-variable and the multi-variable reductions, 
however, we could eventually justify 
the reduction procedure in a unified way, namely, 
by proving that the defining equations (\ref{ddF}) 
of the $F$-function is integrable 
(Theorems \ref{th:integrability} 
and \ref{th:M-integrability}).  
This is a place where the dispersionless Hirota 
equations play a truly fundamental role.  

Viewed from a technical point of view, another clue 
of our method is the use of the generating functions 
(\ref{Faber}) and (\ref{Grunsky}) and 
various identities derived therefrom.  
Not only being closely related to 
the dispersionless Hirota equations themselves, 
these generating functions turned out to be 
also extremely useful in many aspects of 
finite variable reductions.  

Our next target will be, naturally, the cases of 
nonzero genera \cite{Krichever94}.   It will be 
rather easy to derive dispersionless Hirota equations 
for those cases, as partly argued by Krichever et al. 
in a different setup \cite{KMZ05}. 
A main problem is to find a correct form of 
L\"owner-type equations.  We expect to find 
a prototype in differential geometry 
of Hurwitz spaces \cite{Dubrovin-2Dtft,KK01} 
and associated Whitham-type hierarchies 
\cite{Krichever94,Dubrovin92}, because 
they are generalizations of the rational reductions 
that we considered as a prototype of general 
multi-variable reductions for the genus zero case.  
Presumably, we should start with the genus one case, 
for which an explicit description of Hurwitz spaces 
are available in the literature 
\cite{Dubrovin-2Dtft,KS04,RS06,Strachan08} 
along with a candidate of L\"owner-type equations 
\cite{Shramchenko04}.

\subsection*{Acknowledgements}

The authors are partially supported by 
Grant-in-Aid for Scientific Research 
No. 18340061, No. 18540210 and No. 19540179 from 
the Japan Society for the Promotion of Science.

\end{document}